\documentclass[a4paper, 11pt]{article}\usepackage[]{graphicx}\usepackage[]{color}
\makeatletter
\def\maxwidth{ %
  \ifdim\Gin@nat@width>\linewidth
    \linewidth
  \else
    \Gin@nat@width
  \fi
}
\makeatother

\definecolor{fgcolor}{rgb}{0.345, 0.345, 0.345}
\newcommand{\hlnum}[1]{\textcolor[rgb]{0.686,0.059,0.569}{#1}}%
\newcommand{\hlstr}[1]{\textcolor[rgb]{0.192,0.494,0.8}{#1}}%
\newcommand{\hlcom}[1]{\textcolor[rgb]{0.678,0.584,0.686}{\textit{#1}}}%
\newcommand{\hlopt}[1]{\textcolor[rgb]{0,0,0}{#1}}%
\newcommand{\hlstd}[1]{\textcolor[rgb]{0.345,0.345,0.345}{#1}}%
\newcommand{\hlkwb}[1]{\textcolor[rgb]{0.69,0.353,0.396}{#1}}%
\newcommand{\hlkwc}[1]{\textcolor[rgb]{0.333,0.667,0.333}{#1}}%
\newcommand{\hlkwd}[1]{\textcolor[rgb]{0.737,0.353,0.396}{\textbf{#1}}}%

\usepackage{framed}
\makeatletter
\newenvironment{kframe}{%
 \def\at@end@of@kframe{}%
 \ifinner\ifhmode%
  \def\at@end@of@kframe{\end{minipage}}%
  \begin{minipage}{\columnwidth}%
 \fi\fi%
 \def\FrameCommand##1{\hskip\@totalleftmargin \hskip-\fboxsep
 \colorbox{shadecolor}{##1}\hskip-\fboxsep
     \hskip-\linewidth \hskip-\@totalleftmargin \hskip\columnwidth}%
 \MakeFramed {\advance\hsize-\width
   \@totalleftmargin\z@ \linewidth\hsize
   \@setminipage}}%
 {\par\unskip\endMakeFramed%
 \at@end@of@kframe}
\makeatother

\definecolor{shadecolor}{rgb}{.97, .97, .97}
\definecolor{messagecolor}{rgb}{0, 0, 0}
\definecolor{warningcolor}{rgb}{1, 0, 1}
\definecolor{errorcolor}{rgb}{1, 0, 0}
\newenvironment{knitrout}{}{} 

\usepackage{alltt}
\usepackage[english]{babel}
\usepackage[utf8]{inputenc}
\usepackage[T1]{fontenc}
\usepackage{times}
\usepackage{a4wide}
\usepackage{amssymb,amsmath,amsfonts,amsthm}
\usepackage{graphicx}
\usepackage{geometry}
\usepackage{url}
\usepackage[unicode=true,pdfusetitle,
            bookmarks=true,bookmarksnumbered=true,bookmarksopen=true,bookmarksopenlevel=2,
            breaklinks=false,pdfborder={0 0 1},backref=false,colorlinks=true,urlcolor=blue]{hyperref}
\hypersetup{
  pdfstartview={XYZ null null 1}}
\usepackage{breakurl}
\def\be{\begin{equation}}
\def\ee{\end{equation}}
\def\bea{\begin{eqnarray}}
\def\eea{\end{eqnarray}}
\def\lb{\label}
\def\bi{\bibitem}
\def\ct{\cite}

\newcommand{\R}{\textsf{R}}
\newcommand{\leftout}[1]{}
    
\IfFileExists{upquote.sty}{\usepackage{upquote}}{}
\begin{document}
%

%

\title{\sc Tutorial on principal component analysis,\\
with applications in \R{}}
\author{{\sc Henk van Elst}\thanks{ORCID iD:
\href{https://orcid.org/0000-0003-3331-9547}{0000-0003-3331-9547},
E--Mail: $\texttt{Henk.van.Elst@parcIT.de}$} \\
{\small\textit{parcIT GmbH, Erftstraße 15, 50672 K\"{o}ln, Germany}}}

\date{\normalsize{December 6, 2021}}
\maketitle
\sloppy

\begin{abstract}
This tutorial reviews the main steps of the principal component analysis
of a multivariate data set and its subsequent dimensional reduction on the
grounds of identified dominant principal components. The underlying computations
are demonstrated and performed  by means of a script written in the
statistical software package \R{}.

\end{abstract}
%
\begin{flushright}
\textit{In Memoriam: Gisela Ernst (geb.: Freiberg) (1939 to 2020)}
\end{flushright}
%

\section{Introduction}
\lb{sec:intro}
This tutorial demonstrates the main theoretical and practical steps of the 
\textbf{principal component analysis} of a metrically scaled
\textbf{multivariate data set}, and a subsequently performed \textbf{dimensional 
reduction} of the data set considered. The tutorial is based on a transparent 
exposition of all relevant computations by means of a script written with the 
statistical software package \R{} distributed by the R~Core
Team (2020)~\ct{rct2020} free of charge for many different operating systems via
the website~\href{http://cran.r-project.org}{\texttt{cran.r-project.org}}.

\medskip
\noindent
The methodological foundations of the \textbf{principal component analysis} and 
an associated \textbf{dimensional reduction} were laid in particular in the
works by Pearson (1901)~\ct{pea1901}, Hotelling (1933)~\ct{hot1933} and Kaiser
(1960)~\ct{kai1960}; see also Hatzinger \textit{et al}
(2014)~\ct{hatetal2014}, Hair \textit{et al} (2010)~\ct{haietal2010} or Jolliffe 
(2002)~\ct{jol2002}.

\medskip
\noindent
The discussion to follow is divided into three parts. First, in
Secs.~\ref{sec:packages} to \ref{sec:sampladeq}, a \textbf{trivariate example
data set} with \textbf{measured values} for three metrically scaled
\textbf{variables} is loaded and then characterised and visualised with
standard methods of \textbf{Descriptive Statistics}; cf. Ref.~\ct{hve2019}.
Then, in Secs.~\ref{sec:correlmat} to \ref{sec:fmat}, the central tool from
\textbf{Linear Algebra} needed to perfom a \textbf{principal component analysis}
is reviewed: this is the \textbf{eigenvalue analysis} of symmetrical
quadratic matrixes and their \textbf{diagonalisation} by means of
\textbf{rotational transformations} constructed from the matrixes'
\textbf{eigenvectors}. In \textbf{Analytic Geometry} this method is also
known as principal axes transformation; cf. Bronstein \textit{et al}
(2005)~\ct{broetal2005}. Lastly, in Sec.~\ref{sec:dimred}, the procedure of a 
\textbf{dimensional reduction} of a \textbf{multivariate data set} based on an 
\textbf{eigenvalue analysis} of its \textbf{sample correlation matrix} is
outlined in the context of the given \textbf{trivariate example data set}. The 
tutorial ends with a conclusion in Sec.~\ref{sec:concl}.

\medskip
\noindent
The results to be presented have been generated with \R{} Version
4.1.2.

\section{Loading of required \R{} packages}
\lb{sec:packages}
The following \R{} packages and self-written script are loaded into an \R{}
session in order to perform all the calculations involved in this tutorial and
to generate helpful visualisations of the distributions observed in the analysed 
data:
\begin{knitrout}
\definecolor{shadecolor}{rgb}{0.969, 0.969, 0.969}\color{fgcolor}\begin{kframe}
\begin{alltt}
\hlkwd{library}\hlstd{(tidyverse)}
\end{alltt}

{\ttfamily\noindent\itshape\color{messagecolor}{\#\# -- Attaching packages ------------------------------------------------- tidyverse 1.3.1 --}}

{\ttfamily\noindent\itshape\color{messagecolor}{\#\# v ggplot2 3.3.5 \ \ \ \ v purrr \ \ 0.3.4\\\#\# v tibble \ 3.1.6 \ \ \ \ v dplyr \ \ 1.0.7\\\#\# v tidyr \ \ 1.1.4 \ \ \ \ v stringr 1.4.0\\\#\# v readr \ \ 2.1.1 \ \ \ \ v forcats 0.5.1}}

{\ttfamily\noindent\itshape\color{messagecolor}{\#\# -- Conflicts ---------------------------------------------------- tidyverse\_conflicts() --\\\#\# x dplyr::filter() masks stats::filter()\\\#\# x dplyr::lag() \ \ \ masks stats::lag()}}\begin{alltt}
\hlkwd{library}\hlstd{(plotly)}
\end{alltt}

{\ttfamily\noindent\itshape\color{messagecolor}{\#\# \\\#\# Attache Paket: 'plotly'}}

{\ttfamily\noindent\itshape\color{messagecolor}{\#\# Das folgende Objekt ist maskiert 'package:ggplot2':\\\#\# \\\#\# \ \ \ \ last\_plot}}

{\ttfamily\noindent\itshape\color{messagecolor}{\#\# Das folgende Objekt ist maskiert 'package:stats':\\\#\# \\\#\# \ \ \ \ filter}}

{\ttfamily\noindent\itshape\color{messagecolor}{\#\# Das folgende Objekt ist maskiert 'package:graphics':\\\#\# \\\#\# \ \ \ \ layout}}\begin{alltt}
\hlkwd{library}\hlstd{(psych)}
\end{alltt}

{\ttfamily\noindent\itshape\color{messagecolor}{\#\# \\\#\# Attache Paket: 'psych'}}

{\ttfamily\noindent\itshape\color{messagecolor}{\#\# Die folgenden Objekte sind maskiert von 'package:ggplot2':\\\#\# \\\#\# \ \ \ \ \%+\%, alpha}}\begin{alltt}
\hlkwd{library}\hlstd{(REdaS)}
\end{alltt}

{\ttfamily\noindent\itshape\color{messagecolor}{\#\# Lade nötiges Paket: grid}}\begin{alltt}
\hlkwd{library}\hlstd{(e1071)}
\hlkwd{library}\hlstd{(GGally)}
\end{alltt}

{\ttfamily\noindent\itshape\color{messagecolor}{\#\# Registered S3 method overwritten by 'GGally':\\\#\# \ \ method from \ \ \\\#\# \ \ +.gg \ \ ggplot2}}\begin{alltt}
\hlkwd{source}\hlstd{(}\hlkwc{file} \hlstd{=} \hlstr{"descripStats.R"}\hlstd{)}
\end{alltt}
\end{kframe}
\end{knitrout}
%

\section[Data matrix]{Loading of trivariate example data set 
(matrix $\boldsymbol{X}$)}
\lb{sec:data}
The example employed in this tutorial for illustrative purposes is given by
a \textbf{trivariate data set} that contains \textbf{measured values} for the
three metrically scaled \textbf{variables} height~$[\text{cm}]$,
mass~$[\text{kg}]$ and age~$[\text{yr}]$ from a sample of $n = 187$ adult women.
This \textbf{trivariate data set} is part of a larger data set analysed in
Howell (2001)~\ct{how2001}, which needs to be loaded into the \R{}
session.\footnote{The complete original data set may be obtained from the
URL \href{https://tspace.library.utoronto.ca/handle/1807/10395}{tspace.library.utoronto.ca/handle/1807/10395}.}
\begin{knitrout}
\definecolor{shadecolor}{rgb}{0.969, 0.969, 0.969}\color{fgcolor}\begin{kframe}
\begin{alltt}
\hlkwd{load}\hlstd{(}\hlkwc{file} \hlstd{=} \hlstr{"testData.RData"}\hlstd{)}
\hlkwd{str}\hlstd{(}\hlkwc{object} \hlstd{= testData)}
\end{alltt}
\begin{verbatim}
## 'data.frame':	544 obs. of  4 variables:
##  $ height: num  152 140 137 157 145 ...
##  $ weight: num  47.8 36.5 31.9 53 41.3 ...
##  $ age   : num  63 63 65 41 51 35 32 27 19 54 ...
##  $ male  : int  1 0 0 1 0 1 0 1 0 1 ...
\end{verbatim}
\end{kframe}
\end{knitrout}

\medskip
\noindent
The \textbf{trivariate data set} that comprises \textbf{measured values} for the
three \textbf{variables} height~$[\text{cm}]$, mass~$[\text{kg}]$ and
age~$[\text{yr}]$ for a sample of women aged $18~\text{yr}$ or more is obtained
via adequate filtering.
\begin{knitrout}
\definecolor{shadecolor}{rgb}{0.969, 0.969, 0.969}\color{fgcolor}\begin{kframe}
\begin{alltt}
\hlstd{X} \hlkwb{<-}
  \hlstd{testData} \hlopt{%>%}
  \hlstd{dplyr}\hlopt{::}\hlkwd{filter}\hlstd{(}\hlkwc{.data} \hlstd{= ., age} \hlopt{>=} \hlnum{18} \hlopt{&} \hlstd{male} \hlopt{==} \hlnum{0}\hlstd{)} \hlopt{%>%}
  \hlstd{magrittr}\hlopt{::}\hlkwd{set_colnames}\hlstd{(}
    \hlkwc{x} \hlstd{= .,}
    \hlkwc{value} \hlstd{=} \hlkwd{c}\hlstd{(}\hlstr{"height [cm]"}\hlstd{,} \hlstr{"mass [kg]"}\hlstd{,} \hlstr{"age [yr]"}\hlstd{,} \hlstr{"male"}\hlstd{))}
\hlkwd{str}\hlstd{(}\hlkwc{object} \hlstd{= X)}
\end{alltt}
\begin{verbatim}
## 'data.frame':	187 obs. of  4 variables:
##  $ height [cm]: num  140 137 145 149 148 ...
##  $ mass [kg]  : num  36.5 31.9 41.3 38.2 34.9 ...
##  $ age [yr]   : num  63 65 51 32 19 47 73 20 65.3 31 ...
##  $ male       : int  0 0 0 0 0 0 0 0 0 0 ...
\end{verbatim}
\end{kframe}
\end{knitrout}

\medskip
\noindent
The \textbf {data matrix} $\boldsymbol{X}$ constitutes the raw data matrix for
the theoretical and practical considerations outlined in this tutorial.

\subsection{Visualisation of data in $\boldsymbol{X}$ via 3D scatter plot}
\lb{subsec:X3Dscatplot}
To begin with, the \textbf{trivariate data set} in $\boldsymbol{X}$ is first
visualised by means of a \textbf{3D scatter plot}. This is realised by using the
function $\texttt{plot\_ly()}$ from the package $\texttt{plotly}$.\footnote{Note
that the orientation of the scale of measurement along the ``$x$''-axis of this
3D scatter plot does not conform to the mathematical convention; the
resultant reference frame does \textit{not} constitute a right-handed oriented
reference frame.}
\begin{knitrout}
\definecolor{shadecolor}{rgb}{0.969, 0.969, 0.969}\color{fgcolor}\begin{kframe}
\begin{alltt}
\hlstd{fig1} \hlkwb{<-}
  \hlstd{plotly}\hlopt{::}\hlkwd{plot_ly}\hlstd{(}
    \hlkwc{data} \hlstd{= tibble}\hlopt{::}\hlkwd{as_tibble}\hlstd{(}\hlkwc{x} \hlstd{= X),}
    \hlkwc{type} \hlstd{=} \hlstr{"scatter3d"}\hlstd{,}
    \hlkwc{x} \hlstd{= X[,} \hlnum{1}\hlstd{],}
    \hlkwc{y} \hlstd{= X[,} \hlnum{2}\hlstd{],}
    \hlkwc{z} \hlstd{= X[,} \hlnum{3}\hlstd{],}
    \hlkwc{mode} \hlstd{=} \hlstr{"markers"}\hlstd{,}
    \hlkwc{size} \hlstd{=} \hlnum{1}
  \hlstd{)} \hlopt{%>%}
  \hlstd{plotly}\hlopt{::}\hlkwd{layout}\hlstd{(}\hlkwc{title} \hlstd{=} \hlkwd{paste0}\hlstd{(}\hlstr{"Raw data in X "}\hlstd{,}
                                \hlstr{"(original scales of measurement)"}\hlstd{),}
                 \hlkwc{scene} \hlstd{=} \hlkwd{list}\hlstd{(}
                   \hlkwc{xaxis} \hlstd{=} \hlkwd{list}\hlstd{(}\hlkwc{title} \hlstd{=} \hlstr{"height [cm]"}\hlstd{),}
                   \hlkwc{yaxis} \hlstd{=} \hlkwd{list}\hlstd{(}\hlkwc{title} \hlstd{=} \hlstr{"mass [kg]"}\hlstd{),}
                   \hlkwc{zaxis} \hlstd{=} \hlkwd{list}\hlstd{(}\hlkwc{title} \hlstd{=} \hlstr{"age [yr]"}\hlstd{)}
                 \hlstd{))}
\hlstd{fig1}
\end{alltt}
\end{kframe}
\end{knitrout}

\medskip
{\centering \includegraphics[width=14cm]{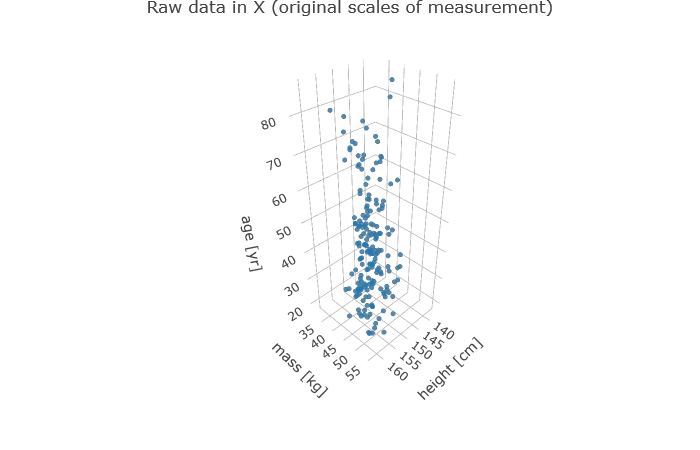}}
%

\subsection{Visualisation of data in $\boldsymbol{X}$ via scatter plot matrix}
\lb{subsec:Xscatplotmat}
In addition, the \textbf{trivariate data set} in $\boldsymbol{X}$ is now
visualised by means of a \textbf{scatter plot matrix}, thus providing
projections of the cloud of data points in the \textbf{3D scatter plot} onto
2D horizontal and vertical slices. For this purpose the function
$\texttt{ggpairs()}$ from the \R{} package $\texttt{GGally}$ is used.
\begin{knitrout}
\definecolor{shadecolor}{rgb}{0.969, 0.969, 0.969}\color{fgcolor}\begin{kframe}
\begin{alltt}
\hlstd{GGally}\hlopt{::}\hlkwd{ggpairs}\hlstd{(}\hlkwc{data} \hlstd{= tibble}\hlopt{::}\hlkwd{as_tibble}\hlstd{(X[,} \hlnum{1}\hlopt{:}\hlnum{3}\hlstd{]))} \hlopt{+}
  \hlkwd{theme_bw}\hlstd{()}
\end{alltt}
\end{kframe}

{\centering \includegraphics[width=\maxwidth]{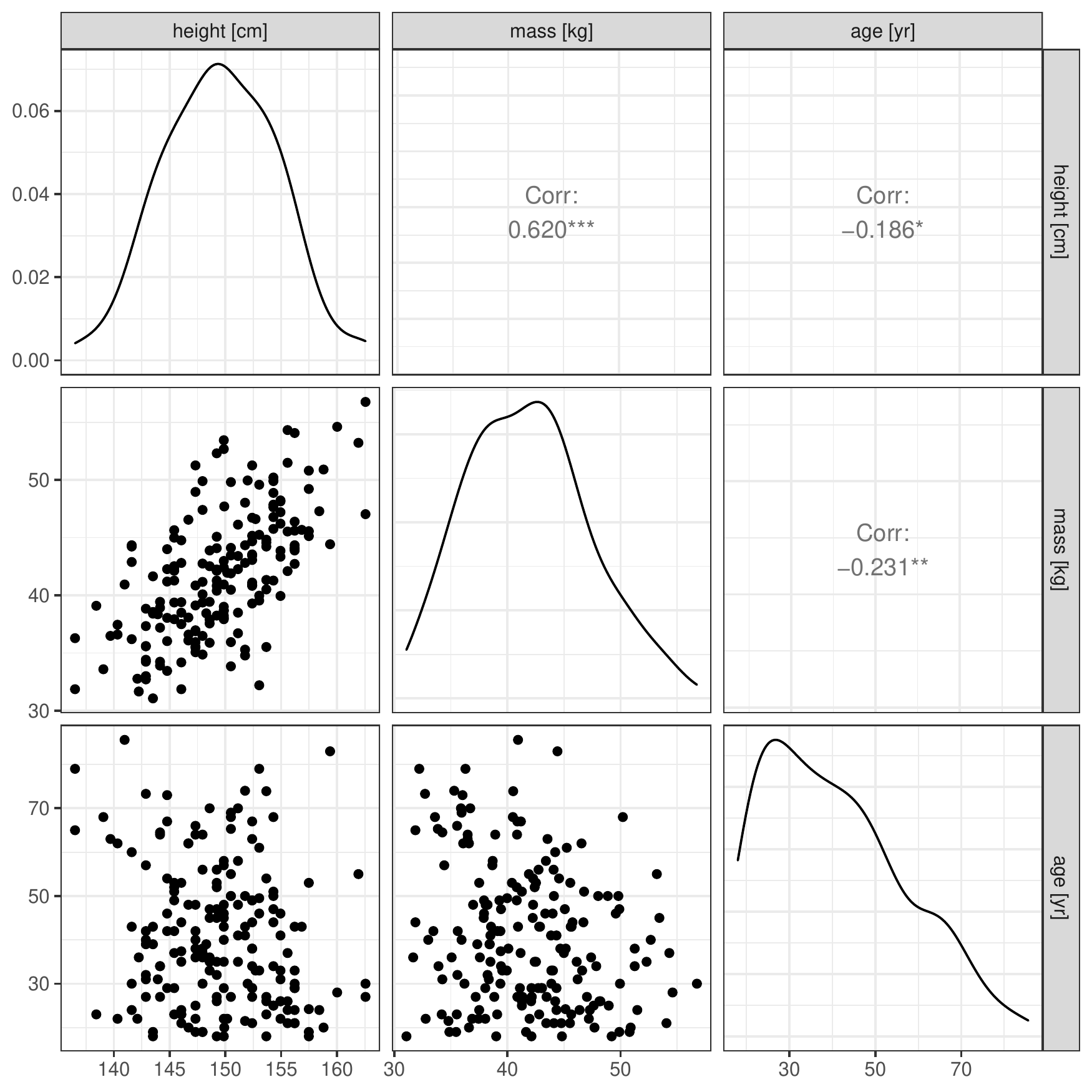} 

}

\end{knitrout}

\medskip
\noindent
Note that the plots of the univariate distributions of the three
\textbf{variables} observed in the present sample, displayed along the diagonal
of the \textbf{scatter plot matrix}, suggest qualitatively that the measured
values for height~$[\text{cm}]$ and mass~$[\text{kg}]$ appear
\textbf{approximately normally distributed}, whereas this property does not
apply to age~$[\text{yr}]$.

\subsection{Descriptive statistics for data in $\boldsymbol{X}$}
\lb{subsec:Xdescrstats}
For the \textbf{measured values} of each of the three \textbf{variables} in
the \textbf{data matrix}~$\boldsymbol{X}$, the following descriptive statistical
measures are computed that characterise their respective observed univariate
distributions:
\begin{enumerate}

\item \textbf{Mean} and \textbf{standard deviation}:
\begin{knitrout}
\definecolor{shadecolor}{rgb}{0.969, 0.969, 0.969}\color{fgcolor}\begin{kframe}
\begin{alltt}
\hlkwd{apply}\hlstd{(}\hlkwc{X} \hlstd{= X[,} \hlnum{1}\hlopt{:}\hlnum{3}\hlstd{],} \hlkwc{MARGIN} \hlstd{=} \hlnum{2}\hlstd{,} \hlkwc{FUN} \hlstd{= mean)}
\end{alltt}
\begin{verbatim}
## height [cm]   mass [kg]    age [yr] 
##   149.51352    41.81419    40.71230
\end{verbatim}
\begin{alltt}
\hlkwd{apply}\hlstd{(}\hlkwc{X} \hlstd{= X[,} \hlnum{1}\hlopt{:}\hlnum{3}\hlstd{],} \hlkwc{MARGIN} \hlstd{=} \hlnum{2}\hlstd{,} \hlkwc{FUN} \hlstd{= sd)}
\end{alltt}
\begin{verbatim}
## height [cm]   mass [kg]    age [yr] 
##    5.084577    5.387917   16.219897
\end{verbatim}
\end{kframe}
\end{knitrout}

\item \textbf{Standardised skewness} and \textbf{standardised excess kurtosis};
cf. Joanes and Gill (1998)~\ct{joagil1998} and van Elst (2019)~\ct{hve2019}:
\begin{knitrout}
\definecolor{shadecolor}{rgb}{0.969, 0.969, 0.969}\color{fgcolor}\begin{kframe}
\begin{alltt}
\hlkwd{standSkewness}\hlstd{(X[,} \hlnum{1}\hlopt{:}\hlnum{3}\hlstd{])}
\end{alltt}
\begin{verbatim}
## height [cm]   mass [kg]    age [yr] 
##   0.0205191   1.7939789   3.3003240
\end{verbatim}
\begin{alltt}
\hlkwd{standKurtosis}\hlstd{(X[,} \hlnum{1}\hlopt{:}\hlnum{3}\hlstd{])}
\end{alltt}
\begin{verbatim}
## height [cm]   mass [kg]    age [yr] 
##  -0.6688236  -0.9719034  -1.3598265
\end{verbatim}
\end{kframe}
\end{knitrout}
As long as the computed values for \textit{both} the standardised skewness and
the standardised excess kurtosis range inside an interval with boundaries
$Q_{0.025} = -1.96$ and $Q_{0.975} = +1.96$, corresponding to the
central $95~\%$ probability interval for a standard normally distributed
variable, one may assume according to statistical convention that one is dealing 
with \textbf{approximately normally distributed} univariate metrically scaled
data; cf. Hair \textit{et al} (2010)~\ct{haietal2010}. As the results obtained
for the \textbf{trivariate example data set} in $\boldsymbol{X}$ show, this
property applies presently to the variables height~$[\text{cm}]$ and
mass~$[\text{kg}]$, but \textit{not} to the variable age~$[\text{yr}]$, the
distribution of which classifies as right-skewed;
cf. the scatter plot matrix discussed in Subsec.~\ref{subsec:Xscatplotmat}.

\item Counts of \textbf{outliers}, \textbf{extremal values} and
\textbf{6-sigma-events}; cf. Toutenburg (2004)~\ct{tou2004}:
\begin{knitrout}
\definecolor{shadecolor}{rgb}{0.969, 0.969, 0.969}\color{fgcolor}\begin{kframe}
\begin{alltt}
\hlkwd{outliers}\hlstd{(X[,} \hlnum{1}\hlopt{:}\hlnum{3}\hlstd{])}
\end{alltt}
\begin{verbatim}
## height [cm]   mass [kg]    age [yr] 
##           0           1           0
\end{verbatim}
\begin{alltt}
\hlkwd{extremalValues}\hlstd{(X[,} \hlnum{1}\hlopt{:}\hlnum{3}\hlstd{])}
\end{alltt}
\begin{verbatim}
## height [cm]   mass [kg]    age [yr] 
##           0           0           0
\end{verbatim}
\begin{alltt}
\hlkwd{sixSigmaEvents}\hlstd{(X[,} \hlnum{1}\hlopt{:}\hlnum{3}\hlstd{])}
\end{alltt}
\begin{verbatim}
## height [cm]   mass [kg]    age [yr] 
##           0           0           0
\end{verbatim}
\end{kframe}
\end{knitrout}

\end{enumerate}
%

\section[Standardised data matrix]{Standardisation of trivariate data set
(matrix $\boldsymbol{Z}$)}
\lb{sec:datastd}
In a next step, the raw data in $\boldsymbol{X}$ needs to be
transformed onto a common \textbf{dimensionless scale of measurement}, with
repect to which \textbf{measured values} for univariate metrically scaled
\textbf{variables} are expressed as \textbf{deviations from the mean in
multiples of the standard deviation}. This kind of \textbf{transformation} is
referred to as \textbf{standardisation}.
\begin{knitrout}
\definecolor{shadecolor}{rgb}{0.969, 0.969, 0.969}\color{fgcolor}\begin{kframe}
\begin{alltt}
\hlstd{Z} \hlkwb{<-}
  \hlkwd{scale}\hlstd{(}\hlkwc{x} \hlstd{= X[,} \hlkwd{c}\hlstd{(}\hlstr{"height [cm]"}\hlstd{,} \hlstr{"mass [kg]"}\hlstd{,} \hlstr{"age [yr]"}\hlstd{)],}
        \hlkwc{center} \hlstd{=} \hlnum{TRUE}\hlstd{,}
        \hlkwc{scale} \hlstd{=} \hlnum{TRUE}\hlstd{)} \hlopt{%>%}
  \hlstd{magrittr}\hlopt{::}\hlkwd{set_colnames}\hlstd{(}
    \hlkwc{x} \hlstd{= .,}
    \hlkwc{value} \hlstd{=} \hlkwd{c}\hlstd{(}\hlstr{"height_std [1]"}\hlstd{,} \hlstr{"mass_std [1]"}\hlstd{,} \hlstr{"age_std [1]"}\hlstd{))}
\hlkwd{dim}\hlstd{(}\hlkwc{x} \hlstd{= Z)}
\end{alltt}
\begin{verbatim}
## [1] 187   3
\end{verbatim}
\begin{alltt}
\hlkwd{head}\hlstd{(}\hlkwc{x} \hlstd{= Z)}
\end{alltt}
\begin{verbatim}
##      height_std [1] mass_std [1] age_std [1]
## [1,]     -1.9300562   -0.9889505   1.3740963
## [2,]     -2.5544936   -1.8466045   1.4974016
## [3,]     -0.8060688   -0.0997265   0.6342642
## [4,]     -0.0567439   -0.6627263  -0.5371365
## [5,]     -0.3065189   -1.2888663  -1.3386213
## [6,]      0.9423560    1.4998244   0.3876535
\end{verbatim}
\end{kframe}
\end{knitrout}

\medskip
\noindent
In consequence of \textbf{standardisation}, the univariate data for each of the
three \textbf{variables} in the resultant \textbf{data matrix}~$\boldsymbol{Z}$
exhibit a \textbf{mean} of $0$ and a \textbf{standard deviation} of $1$. This
property will be displayed shortly.

\medskip
\noindent
The \textbf{data matrix}~$\boldsymbol{Z}$ of \textbf{standardised 
measured values} (``$z$-scores'') constitutes the basis of all subsequent
steps of \textbf{statistical data analysis}.

\subsection{Visualisation of data in $\boldsymbol{Z}$ via 3D scatter plot}
\lb{subsec:Z3Dscatplot}
%
\begin{knitrout}
\definecolor{shadecolor}{rgb}{0.969, 0.969, 0.969}\color{fgcolor}\begin{kframe}
\begin{alltt}
\hlstd{fig2} \hlkwb{<-}
  \hlstd{plotly}\hlopt{::}\hlkwd{plot_ly}\hlstd{(}
    \hlkwc{data} \hlstd{= tibble}\hlopt{::}\hlkwd{as_tibble}\hlstd{(}\hlkwc{x} \hlstd{= Z),}
    \hlkwc{type} \hlstd{=} \hlstr{"scatter3d"}\hlstd{,}
    \hlkwc{x} \hlstd{= Z[,} \hlnum{1}\hlstd{],}
    \hlkwc{y} \hlstd{= Z[,} \hlnum{2}\hlstd{],}
    \hlkwc{z} \hlstd{= Z[,} \hlnum{3}\hlstd{],}
    \hlkwc{mode} \hlstd{=} \hlstr{"markers"}\hlstd{,}
    \hlkwc{size} \hlstd{=} \hlnum{1}
  \hlstd{)} \hlopt{%>%}
  \hlstd{plotly}\hlopt{::}\hlkwd{layout}\hlstd{(}\hlkwc{title} \hlstd{=} \hlkwd{paste0}\hlstd{(}\hlstr{"Standardised data in Z "}\hlstd{,}
                                \hlstr{"(unit scale of measurement)"}\hlstd{),}
                 \hlkwc{scene} \hlstd{=} \hlkwd{list}\hlstd{(}
                   \hlkwc{xaxis} \hlstd{=} \hlkwd{list}\hlstd{(}\hlkwc{title} \hlstd{=} \hlstr{"height std [1]"}\hlstd{),}
                   \hlkwc{yaxis} \hlstd{=} \hlkwd{list}\hlstd{(}\hlkwc{title} \hlstd{=} \hlstr{"mass std [1]"}\hlstd{),}
                   \hlkwc{zaxis} \hlstd{=} \hlkwd{list}\hlstd{(}\hlkwc{title} \hlstd{=} \hlstr{"age std [1]"}\hlstd{)}
                 \hlstd{))}
\hlstd{fig2}
\end{alltt}
\end{kframe}
\end{knitrout}

\medskip
{\centering \includegraphics[width=14cm]{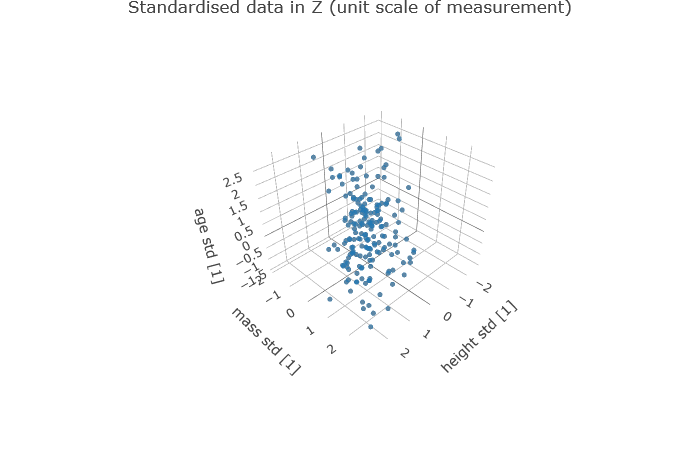}}
%

\subsection{Visualisation of data in $\boldsymbol{Z}$ via scatter plot matrix}
\lb{subsec:Zscatplotmat}
%
\begin{knitrout}
\definecolor{shadecolor}{rgb}{0.969, 0.969, 0.969}\color{fgcolor}\begin{kframe}
\begin{alltt}
\hlstd{GGally}\hlopt{::}\hlkwd{ggpairs}\hlstd{(}\hlkwc{data} \hlstd{= tibble}\hlopt{::}\hlkwd{as_tibble}\hlstd{(}\hlkwc{x} \hlstd{= Z))} \hlopt{+}
  \hlkwd{theme_bw}\hlstd{()}
\end{alltt}
\end{kframe}

{\centering \includegraphics[width=\maxwidth]{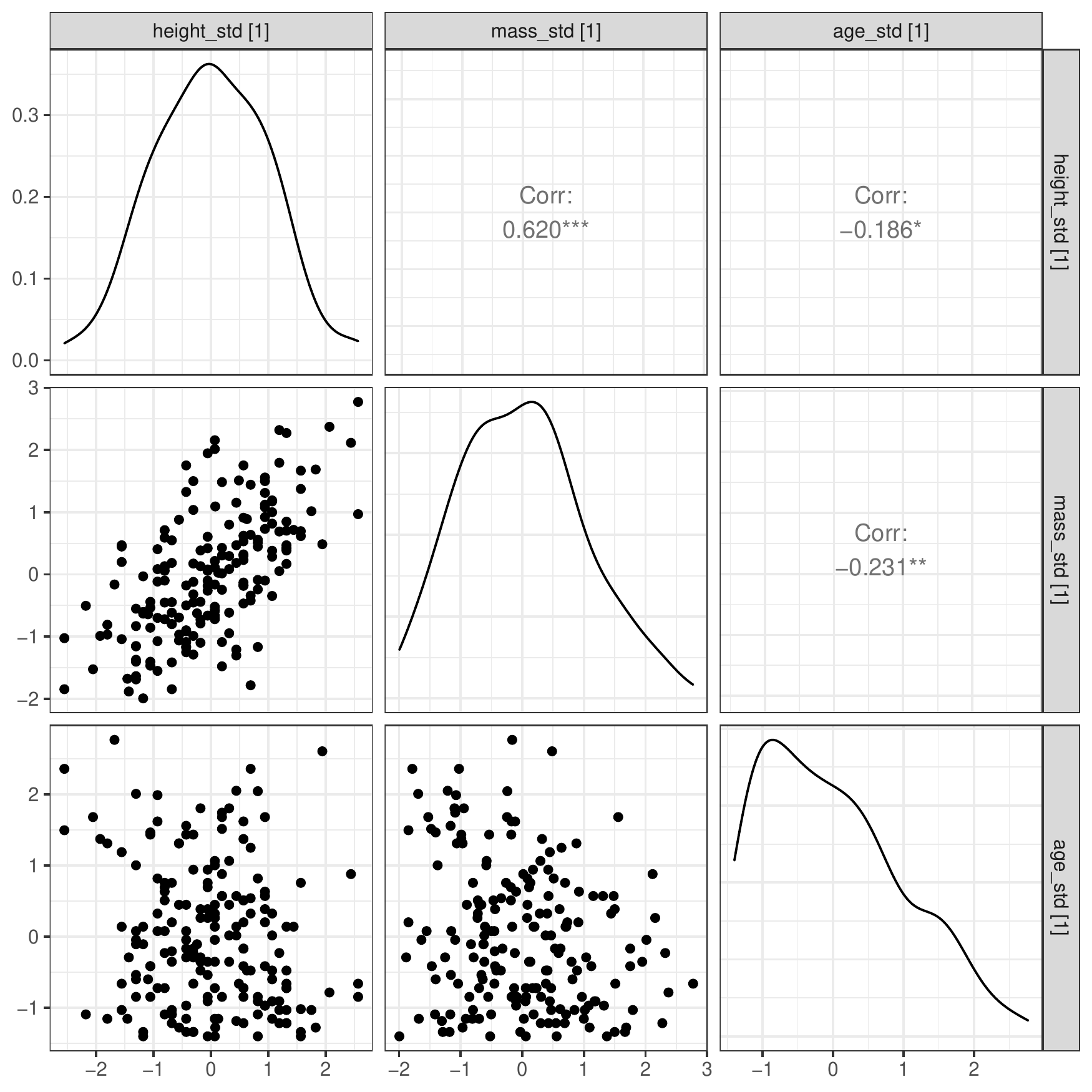} 

}

\end{knitrout}

\medskip
\noindent
As the comparison of this \textbf{scatter plot matrix} with the one given in
Subsec.~\ref{subsec:Zscatplotmat} for the original \textbf{measured values}
shows qualitatively, \textbf{standardisation} has preserved the observed
uni- and bivariate (and trivariate) distributional properties of the
\textbf{trivariate data set}. Otherwise, a \textbf{transformation} of this 
type would have to be rejected as an illegitimate procedure.

\subsection{Visualisation of data in $\boldsymbol{Z}$ via parallel box plots}
\lb{subsec:Zboxplots}
%
\begin{knitrout}
\definecolor{shadecolor}{rgb}{0.969, 0.969, 0.969}\color{fgcolor}\begin{kframe}
\begin{alltt}
\hlstd{Z} \hlopt{%>%}
  \hlstd{tibble}\hlopt{::}\hlkwd{as_tibble}\hlstd{(}\hlkwc{x} \hlstd{= .)} \hlopt{%>%}
  \hlstd{tidyr}\hlopt{::}\hlkwd{pivot_longer}\hlstd{(}
    \hlkwc{data} \hlstd{= .,}
    \hlkwc{cols} \hlstd{=} \hlkwd{c}\hlstd{(`height_std [1]`, `mass_std [1]`, `age_std [1]`),}
    \hlkwc{names_to} \hlstd{=} \hlstr{"dimension"}\hlstd{,}
    \hlkwc{values_to} \hlstd{=} \hlstr{"zscore"}
  \hlstd{)} \hlopt{%>%}
  \hlstd{dplyr}\hlopt{::}\hlkwd{mutate}\hlstd{(}
    \hlkwc{dimension} \hlstd{= forcats}\hlopt{::}\hlkwd{fct_relevel}\hlstd{(}
      \hlkwc{.f} \hlstd{= dimension,}
      \hlkwd{c}\hlstd{(}\hlstr{"height_std [1]"}\hlstd{,} \hlstr{"mass_std [1]"}\hlstd{,} \hlstr{"age_std [1]"}\hlstd{))}
    \hlstd{)} \hlopt{%>%}
  \hlkwd{ggplot}\hlstd{(}\hlkwc{data} \hlstd{= .,} \hlkwc{mapping} \hlstd{=} \hlkwd{aes}\hlstd{(}\hlkwc{x} \hlstd{= dimension,} \hlkwc{y} \hlstd{= zscore))} \hlopt{+}
  \hlkwd{geom_boxplot}\hlstd{()} \hlopt{+}
  \hlkwd{xlab}\hlstd{(}\hlkwc{label} \hlstd{=} \hlstr{"dimension"}\hlstd{)} \hlopt{+}
  \hlkwd{ylab}\hlstd{(}\hlkwc{label} \hlstd{=} \hlstr{"zscore [1]"}\hlstd{)} \hlopt{+}
  \hlkwd{theme_bw}\hlstd{()}
\end{alltt}
\end{kframe}

{\centering \includegraphics[width=\maxwidth]{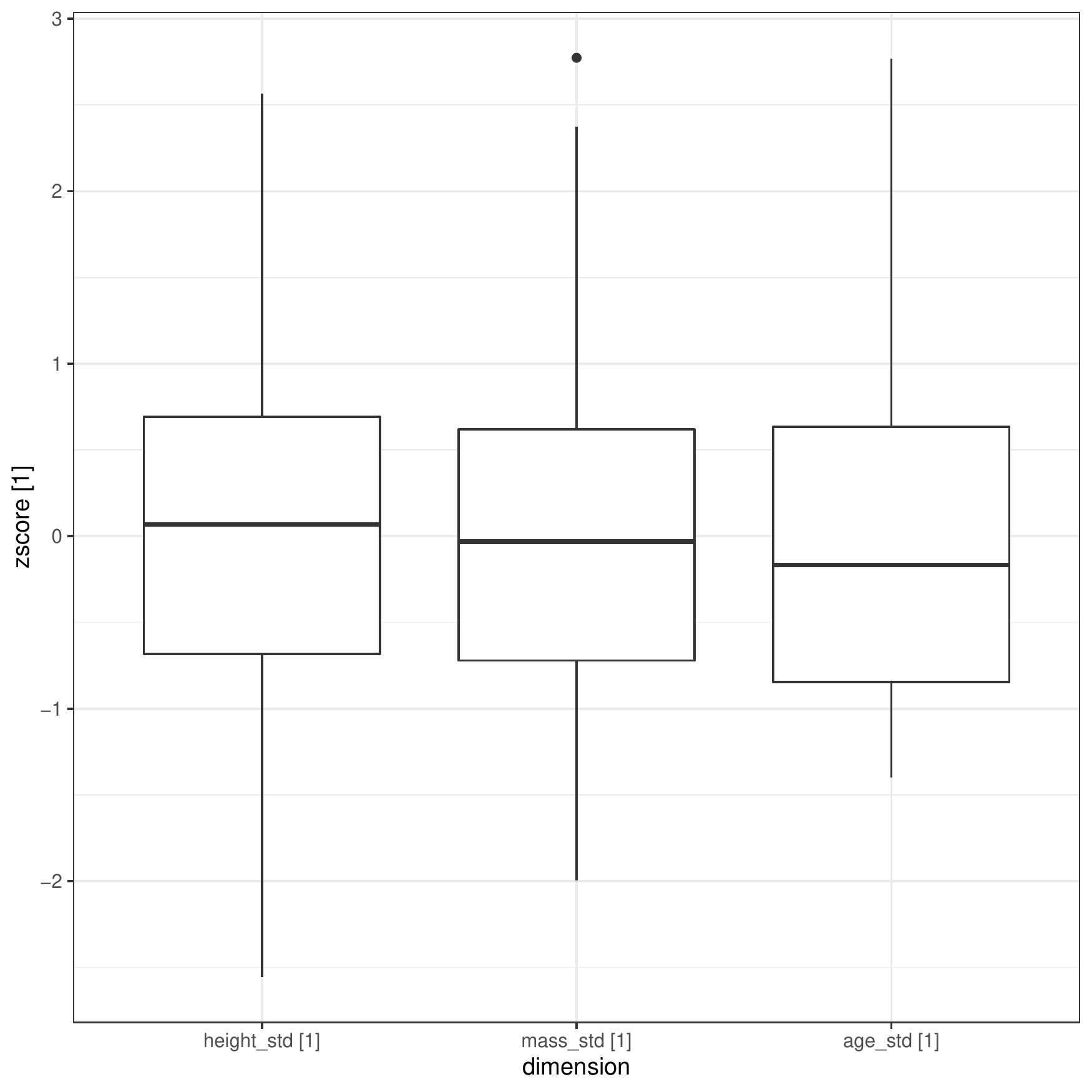} 

}

\end{knitrout}

\medskip
\noindent
The \textbf{parallel box plots} point to the existence of a
\textbf{heterogeneity of variances} between the univariate data for each of
the three variables in the \textbf{data matrix}~$\boldsymbol{Z}$.

\medskip
\noindent
The preservation of the distributional properties of the data under
\textbf{standardisation} is demonstrated from the univariate perspective by the
next step of \textbf{statistical data analysis}.

\subsection{Descriptive statistics for data in $\boldsymbol{Z}$}
\lb{subsec:Zdescrstats}
Again, descriptive statistical measures are computed that characterise observed
univariate distributions, this time for the data in $\boldsymbol{Z}$
(``$z$-scores''):
\begin{enumerate}

\item \textbf{Mean} and \textbf{standard deviation}:
\begin{knitrout}
\definecolor{shadecolor}{rgb}{0.969, 0.969, 0.969}\color{fgcolor}\begin{kframe}
\begin{alltt}
\hlkwd{apply}\hlstd{(}\hlkwc{X} \hlstd{= Z,} \hlkwc{MARGIN} \hlstd{=} \hlnum{2}\hlstd{,} \hlkwc{FUN} \hlstd{= mean)} \hlopt{%>%}
  \hlkwd{round}\hlstd{(}\hlkwc{x} \hlstd{= .,} \hlkwc{digits} \hlstd{=} \hlnum{4}\hlstd{)}
\end{alltt}
\begin{verbatim}
## height_std [1]   mass_std [1]    age_std [1] 
##              0              0              0
\end{verbatim}
\begin{alltt}
\hlkwd{apply}\hlstd{(}\hlkwc{X} \hlstd{= Z,} \hlkwc{MARGIN} \hlstd{=} \hlnum{2}\hlstd{,} \hlkwc{FUN} \hlstd{= sd)}
\end{alltt}
\begin{verbatim}
## height_std [1]   mass_std [1]    age_std [1] 
##              1              1              1
\end{verbatim}
\end{kframe}
\end{knitrout}

\item \textbf{Standardised skewness} and \textbf{standardised excess kurtosis};
cf. Joanes and Gill (1998)~\ct{joagil1998} and van Elst (2019)~\ct{hve2019}:
\begin{knitrout}
\definecolor{shadecolor}{rgb}{0.969, 0.969, 0.969}\color{fgcolor}\begin{kframe}
\begin{alltt}
\hlkwd{standSkewness}\hlstd{(Z)}
\end{alltt}
\begin{verbatim}
## height_std [1]   mass_std [1]    age_std [1] 
##      0.0205191      1.7939789      3.3003240
\end{verbatim}
\begin{alltt}
\hlkwd{standKurtosis}\hlstd{(Z)}
\end{alltt}
\begin{verbatim}
## height_std [1]   mass_std [1]    age_std [1] 
##     -0.6688236     -0.9719034     -1.3598265
\end{verbatim}
\end{kframe}
\end{knitrout}

\end{enumerate}
%

\section[Sampling adequacy for principal component analysis]{Sampling adequacy
of trivariate data set in $\boldsymbol{Z}$ for principal component analysis}
\lb{sec:sampladeq}
The \textbf{sampling adequacy} for a \textbf{principal component analysis} of
the present \textbf{trivariate data set} is evaluated by employing Bartlett's
(1951)~\ct{bar1951} \textbf{test of sphericity}, as well as the standardised
\textbf{KMO} and \textbf{MSA measures} according to Kaiser, Meyer und Olkin
(KMO); cf. Kaiser (1970)~\ct{kai1960}, Guttman (1953)~\ct{gut1953} and 
Hatzinger \textit{et al} (2014)~\ct{hatetal2014}. For this purpose, one may use
the functions $\texttt{bart\_spher()}$ and $\texttt{KMO()}$ from the \R{}
package $\texttt{REdaS}$.

\medskip
\noindent
\textbf{Bartlett's} (1951)~\ct{bar1951} frequentist \textbf{null hypothesis
significance test} subjects the assumption of sphericity of the
\textbf{envelope} of the \textbf{cloud of data points} (defined via the
matrix~$\boldsymbol{Z}$) in, presently, Euclidian space~$\mathbb{R}^{3}$ to an
empirical check. For the given \textbf{trivariate data set} this yields
\begin{knitrout}
\definecolor{shadecolor}{rgb}{0.969, 0.969, 0.969}\color{fgcolor}\begin{kframe}
\begin{alltt}
\hlstd{REdaS}\hlopt{::}\hlkwd{bart_spher}\hlstd{(}\hlkwc{x} \hlstd{= Z)}
\end{alltt}
\begin{verbatim}
## 	Bartlett's Test of Sphericity
## 
## Call: REdaS::bart_spher(x = Z)
## 
##      X2 = 100.12
##      df = 3
## p-value < 2.22e-16
\end{verbatim}
\end{kframe}
\end{knitrout}

\medskip
\noindent
In view of the calculated $p$-value, the null hypothesis can be rejected at a
significance level of $\alpha = 0.01$. Most likely the empirically attested
deformation of the \textbf{envelope} is not due to chance. Non-sphericity of
the \textbf{envelope} may thus be assumed and so supports the intention of
performing a \textbf{principal component analysis} on the \textbf{trivariate
data set} in $\boldsymbol{Z}$.\footnote{The non-sphericity of the envelope of
the cloud of data points defined via~$\boldsymbol{Z}$ is conspicuous in the 3D
scatter plot displayed in Subsec.~\ref{subsec:Z3Dscatplot}.}

\medskip
\noindent
The standardised \textbf{KMO} and \textbf{MSA measures}, which both take values
in the interval $[0; 1]$, assume for the \textbf{trivariate data set} in
$\boldsymbol{Z}$ the values
\begin{knitrout}
\definecolor{shadecolor}{rgb}{0.969, 0.969, 0.969}\color{fgcolor}\begin{kframe}
\begin{alltt}
\hlstd{kmoZ} \hlkwb{<-}
  \hlstd{REdaS}\hlopt{::}\hlkwd{KMOS}\hlstd{(}\hlkwc{x} \hlstd{= Z)}
\hlkwd{print}\hlstd{(}\hlkwc{x} \hlstd{= kmoZ,} \hlkwc{stats} \hlstd{=} \hlstr{"KMO"}\hlstd{)}
\end{alltt}
\begin{verbatim}
## 
## Kaiser-Meyer-Olkin Statistic
## Call: REdaS::KMOS(x = Z)
## 
## KMO-Criterion: 0.5478232
\end{verbatim}
\begin{alltt}
\hlkwd{print}\hlstd{(}
  \hlkwc{x} \hlstd{= kmoZ,}
  \hlkwc{stats} \hlstd{=} \hlstr{"MSA"}\hlstd{,}
  \hlkwc{sort} \hlstd{=} \hlnum{TRUE}\hlstd{,}
  \hlkwc{digits} \hlstd{=} \hlnum{7}\hlstd{,}
  \hlkwc{show} \hlstd{=} \hlnum{1}\hlopt{:}\hlnum{3}
\hlstd{)}
\end{alltt}
\begin{verbatim}
## 
## Kaiser-Meyer-Olkin Statistics
## 
## Call: REdaS::KMOS(x = Z)
## 
## Measures of Sampling Adequacy (MSA):
##   mass_std [1] height_std [1]    age_std [1] 
##      0.5309161      0.5327821      0.7749174
\end{verbatim}
\end{kframe}
\end{knitrout}

\medskip
\noindent
The \textbf{KMO measure} quantifies the \textbf{sampling adequacy} of the entire 
\textbf{data set} in $\boldsymbol{Z}$, whereas, in contrast, the \textbf{MSA
measure} individually quantifies the \textbf{sampling adequacy} of the
\textbf{measured values} for every single \textbf{variable}. According to, e.g.,
Hatzinger \textit{et al} (2014) ~\ct{hatetal2014} or Hair \textit{et al}
(2010)~\ct{haietal2010}, a good \textbf{sampling adequacy} for the \textbf{data
set} in $\boldsymbol{Z}$ is given when \textit{both} standardised measures
range between $0.8$ und $1.0$. Of course, in this respect, the presently
considered \textbf{trivariate data set} in $\boldsymbol{Z}$ constitutes a
\textit{negative example}. However, as it proves very useful for demonstrating 
the main theoretical and practical steps of the \textbf{principal component
analysis} of a metrically scaled \textbf{multivariate data set}, and is also
accessible to the reader's imagination, this tutorial continuous to employ it in 
the steps of \textbf{statistical data analysis} taken in the following sections.

\section[Sample correlation matrix and its inverse]{Calculation of sample
correlation matrix $\boldsymbol{R}$ and its inverse $\boldsymbol{R}^{-1}$}
\lb{sec:correlmat}
The \textbf{sample correlation matrix}~$\boldsymbol{R}$ of the considered
\textbf{trivariate data set} in $\boldsymbol{X}$ is defined in terms of
algebraic projections of ``$z$-scores'' onto themselves by
\[
{\displaystyle \boldsymbol{R} := \frac{1}{n-1}\,\boldsymbol{Z}^{\top}
\boldsymbol{Z}} \ ,
\]
viz.
\begin{knitrout}
\definecolor{shadecolor}{rgb}{0.969, 0.969, 0.969}\color{fgcolor}\begin{kframe}
\begin{alltt}
\hlstd{Rmat} \hlkwb{<-}
  \hlstd{(}\hlnum{1} \hlopt{/} \hlstd{(}\hlkwd{nrow}\hlstd{(Z)} \hlopt{-} \hlnum{1}\hlstd{))} \hlopt{*} \hlkwd{t}\hlstd{(Z)} \hlopt{%*%} \hlstd{Z}
\hlstd{Rmat}
\end{alltt}
\begin{verbatim}
##                height_std [1] mass_std [1] age_std [1]
## height_std [1]      1.0000000    0.6202596  -0.1863417
## mass_std [1]        0.6202596    1.0000000  -0.2308225
## age_std [1]        -0.1863417   -0.2308225   1.0000000
\end{verbatim}
\end{kframe}
\end{knitrout}

\medskip
\noindent
The \textbf{sample correlation matrix}~$\boldsymbol{R}$ exhibits a non-zero
value for its \textbf{determinant} and therefore classifies as regular. It
follows that there exists an \textbf{inverse}, $\boldsymbol{R}^{-1}$, which is
given here for completeness:
\begin{knitrout}
\definecolor{shadecolor}{rgb}{0.969, 0.969, 0.969}\color{fgcolor}\begin{kframe}
\begin{alltt}
\hlkwd{det}\hlstd{(}\hlkwc{x} \hlstd{= Rmat)}
\end{alltt}
\begin{verbatim}
## [1] 0.5806328
\end{verbatim}
\begin{alltt}
\hlstd{RmatInv} \hlkwb{<-}
  \hlkwd{solve}\hlstd{(Rmat)}
\hlstd{RmatInv}
\end{alltt}
\begin{verbatim}
##                height_std [1] mass_std [1] age_std [1]
## height_std [1]     1.63049873   -0.9941702  0.07435314
## mass_std [1]      -0.99417018    1.6624566  0.19847692
## age_std [1]        0.07435314    0.1984769  1.05966802
\end{verbatim}
\end{kframe}
\end{knitrout}

\medskip
\noindent
For the example considered, the \textbf{trace} of the \textbf{sample 
correlation matrix}~$\boldsymbol{R}$ amounts to
\begin{knitrout}
\definecolor{shadecolor}{rgb}{0.969, 0.969, 0.969}\color{fgcolor}\begin{kframe}
\begin{alltt}
\hlkwd{sum}\hlstd{(}\hlkwd{diag}\hlstd{(}\hlkwc{x} \hlstd{= Rmat))}
\end{alltt}
\begin{verbatim}
## [1] 3
\end{verbatim}
\end{kframe}
\end{knitrout}

\medskip
\noindent
This value is equal to the number of \textbf{variables} in the considered
\textbf{trivariate data set} in $\boldsymbol{X}$.

\section[Eigenvalues and eigenvectors]{Eigenvalues and eigenvectors:
orthonormal eigenbasis of sample correlation matrix}
\lb{sec:eigen}
In the present case, the three \textbf{eigenvalues} of the \textbf{sample
correlation matrix}~$\boldsymbol{R}$ are
\begin{knitrout}
\definecolor{shadecolor}{rgb}{0.969, 0.969, 0.969}\color{fgcolor}\begin{kframe}
\begin{alltt}
\hlstd{evAnaCor} \hlkwb{<-}
  \hlkwd{eigen}\hlstd{(}\hlkwc{x} \hlstd{= Rmat,} \hlkwc{symmetric} \hlstd{=} \hlstr{"TRUE"}\hlstd{)}
\hlstd{evAnaCor}\hlopt{$}\hlstd{values}
\end{alltt}
\begin{verbatim}
## [1] 1.7382412 0.8838105 0.3779482
\end{verbatim}
\begin{alltt}
\hlkwd{sum}\hlstd{(evAnaCor}\hlopt{$}\hlstd{values)}
\end{alltt}
\begin{verbatim}
## [1] 3
\end{verbatim}
\end{kframe}
\end{knitrout}

\medskip
\noindent
and they sum up to the very number of \textbf{variables} in the 
\textbf{trivariate data set} in $\boldsymbol{X}$. The three mutually orthogonal,
normalised \textbf{eigenvectors} of the \textbf{sample correlation
matrix}~$\boldsymbol{R}$ are (columnwise, from left to
right)\footnote{Regrettably \R{} does not return the components of the three
identified eigenvectors according to the usual mathematical convention, i.e., so
that the eigenvectors form a right-handed oriented orthonormal basis.}
\begin{knitrout}
\definecolor{shadecolor}{rgb}{0.969, 0.969, 0.969}\color{fgcolor}\begin{kframe}
\begin{alltt}
\hlstd{evAnaCor}\hlopt{$}\hlstd{vectors}
\end{alltt}
\begin{verbatim}
##            [,1]      [,2]        [,3]
## [1,] -0.6504363 0.3033698  0.69634715
## [2,] -0.6625952 0.2215906 -0.71544756
## [3,]  0.3713492 0.9267494 -0.05688085
\end{verbatim}
\end{kframe}
\end{knitrout}

\medskip
\noindent
These are also referred to as the \textbf{principal components} of the
\textbf{trivariate data set} in ~$\boldsymbol{X}$. The \textbf{eigenvectors}
span the \textbf{orthonormal eigenbasis} of the \textbf{sample correlation
matrix}~$\boldsymbol{R}$ in Euclidian space~$\mathbb{R}^{3}$; cf. Bronstein 
\textit{et al} (2005)~\ct{broetal2005}.

\medskip
\noindent
With regard to Kaiser's (1960)~\ct{kai1960} \textbf{eigenvalue criterion} it is
noted that in the present example only one of the three \textbf{eigenvalues} of 
the \textbf{sample correlation matrix}~$\boldsymbol{R}$ is greater than $1$,
implying that presently the \textbf{trivariate data set} in $\boldsymbol{X}$ 
possesses only a single \textbf{dominant principal component}.

\medskip
\noindent
The \textbf{proportions of total variance} of the \textbf{trivariate data set}
in $\boldsymbol{Z}$ explained by each of the three \textbf{principal 
components} amount to
\begin{knitrout}
\definecolor{shadecolor}{rgb}{0.969, 0.969, 0.969}\color{fgcolor}\begin{kframe}
\begin{alltt}
\hlkwd{round}\hlstd{(}\hlkwc{x} \hlstd{= (evAnaCor}\hlopt{$}\hlstd{values} \hlopt{/} \hlkwd{sum}\hlstd{(evAnaCor}\hlopt{$}\hlstd{values)),} \hlkwc{digits} \hlstd{=} \hlnum{4}\hlstd{)}
\end{alltt}
\begin{verbatim}
## [1] 0.5794 0.2946 0.1260
\end{verbatim}
\end{kframe}
\end{knitrout}

\medskip
\noindent
i.e., $57.94~\%$, $29.46~\%$ and $12.60~\%$, respectively, and in cumulative
terms
\begin{knitrout}
\definecolor{shadecolor}{rgb}{0.969, 0.969, 0.969}\color{fgcolor}\begin{kframe}
\begin{alltt}
\hlkwd{round}\hlstd{(}\hlkwc{x} \hlstd{= (}\hlkwd{cumsum}\hlstd{(evAnaCor}\hlopt{$}\hlstd{values)} \hlopt{/} \hlkwd{sum}\hlstd{(evAnaCor}\hlopt{$}\hlstd{values)),} \hlkwc{digits} \hlstd{=} \hlnum{4}\hlstd{)}
\end{alltt}
\begin{verbatim}
## [1] 0.5794 0.8740 1.0000
\end{verbatim}
\end{kframe}
\end{knitrout}

\medskip
\noindent
An interpretation for the \textbf{eigenvalues} of the \textbf{sample correlation 
matrix}~$\boldsymbol{R}$ will be given in the next section.

\section[Rotation matrix and diagonal eigenvalue matrix]{Rotation matrix
$\boldsymbol{V}$, diagonal eigenvalue matrix $\boldsymbol{\Lambda}$ and inverse 
diagonal eigenvalue matrix $\boldsymbol{\Lambda}^{-1}$}
\lb{sec:rotmat}
From the three eigenvectors of the \textbf{sample correlation
matrix}~$\boldsymbol{R}$ one constructs an orthogonal \textbf{rotation 
matrix}~$\boldsymbol{V}$, by means of which one can perform (presently in
Euclidian space~$\mathbb{R}^{3}$) transformations to the \textit{right-handed
oriented} \textbf{orthonormal eigenbasis} of the \textbf{sample correlation
matrix}~$\boldsymbol{R}$. The \textbf{determinant} of the \textbf{rotation
matrix}~$\boldsymbol{V}$ has the value~$1$, i.e.,
\begin{knitrout}
\definecolor{shadecolor}{rgb}{0.969, 0.969, 0.969}\color{fgcolor}\begin{kframe}
\begin{alltt}
\hlstd{rotMatCor} \hlkwb{<-}
  \hlstd{evAnaCor}\hlopt{$}\hlstd{vectors}
\hlstd{rotMatCor[,} \hlnum{3}\hlstd{]} \hlkwb{<-}
  \hlstd{(}\hlopt{-}\hlnum{1}\hlstd{)} \hlopt{*} \hlstd{rotMatCor[,} \hlnum{3}\hlstd{]}  \hlcom{# re-scaling by a factor of (-1)}
\hlstd{rotMatCor}
\end{alltt}
\begin{verbatim}
##            [,1]      [,2]        [,3]
## [1,] -0.6504363 0.3033698 -0.69634715
## [2,] -0.6625952 0.2215906  0.71544756
## [3,]  0.3713492 0.9267494  0.05688085
\end{verbatim}
\begin{alltt}
\hlkwd{det}\hlstd{(}\hlkwc{x} \hlstd{= rotMatCor)}
\end{alltt}
\begin{verbatim}
## [1] 1
\end{verbatim}
\end{kframe}
\end{knitrout}

\medskip
\noindent
implying that transformations with the \textbf{rotation matrix}~$\boldsymbol{V}$
\textit{preserve volumes}.

\medskip
\noindent
Per construction the \textbf{rotation matrix}~$\boldsymbol{V}$ satisfies the
following two \textbf{tests of orthogonality}
\[
\boldsymbol{1} = \boldsymbol{V}^{\top}
\boldsymbol{V} = \boldsymbol{V}\boldsymbol{V}^{\top} \ ,
\]
viz.
\begin{knitrout}
\definecolor{shadecolor}{rgb}{0.969, 0.969, 0.969}\color{fgcolor}\begin{kframe}
\begin{alltt}
\hlkwd{round}\hlstd{(}\hlkwc{x} \hlstd{=} \hlkwd{t}\hlstd{(rotMatCor)} \hlopt{%*%} \hlstd{rotMatCor,} \hlkwc{digits} \hlstd{=} \hlnum{4}\hlstd{)}
\end{alltt}
\begin{verbatim}
##      [,1] [,2] [,3]
## [1,]    1    0    0
## [2,]    0    1    0
## [3,]    0    0    1
\end{verbatim}
\begin{alltt}
\hlkwd{round}\hlstd{(}\hlkwc{x} \hlstd{= rotMatCor} \hlopt{%*%} \hlkwd{t}\hlstd{(rotMatCor),} \hlkwc{digits} \hlstd{=} \hlnum{4}\hlstd{)}
\end{alltt}
\begin{verbatim}
##      [,1] [,2] [,3]
## [1,]    1    0    0
## [2,]    0    1    0
## [3,]    0    0    1
\end{verbatim}
\end{kframe}
\end{knitrout}

\medskip
\noindent
By means of \textbf{diagonalisation} of the \textbf{sample correlation 
matrix}~$\boldsymbol{R}$ via transformation with the \textbf{rotation 
matrix}~$\boldsymbol{V}$, one obtains the \textbf{diagonal eigenvalue 
matrix}~$\boldsymbol{\Lambda}$ as
\[
\boldsymbol{\Lambda} = \boldsymbol{V}^{\top}\boldsymbol{R}\boldsymbol{V} \ ,
\]
viz.
\begin{knitrout}
\definecolor{shadecolor}{rgb}{0.969, 0.969, 0.969}\color{fgcolor}\begin{kframe}
\begin{alltt}
\hlstd{LambdaCor} \hlkwb{<-}
  \hlkwd{t}\hlstd{(rotMatCor)} \hlopt{%*%} \hlstd{Rmat} \hlopt{%*%} \hlstd{rotMatCor}
\hlkwd{round}\hlstd{(}\hlkwc{x} \hlstd{= LambdaCor,} \hlkwc{digits} \hlstd{=} \hlnum{7}\hlstd{)}
\end{alltt}
\begin{verbatim}
##          [,1]      [,2]      [,3]
## [1,] 1.738241 0.0000000 0.0000000
## [2,] 0.000000 0.8838105 0.0000000
## [3,] 0.000000 0.0000000 0.3779482
\end{verbatim}
\end{kframe}
\end{knitrout}

\medskip
\noindent
The \textbf{diagonal eigenvalue matrix}~$\boldsymbol{\Lambda}$ is nothing but
the representation of the \textbf{sample correlation matrix}~$\boldsymbol{R}$
with respect to its \textbf{orthonormal eigenbasis} in Euclidian
space~$\mathbb{R}^{3}$.

\medskip
\noindent
The \textbf{diagonal eigenvalue matrix}~$\boldsymbol{\Lambda}$ obtained above
may be subjected to the \textbf{consistency check}
\[
\textbf{0} = \boldsymbol{\Lambda} -
\text{diag}\left(\lambda_{1}, \ldots, \lambda_{m}\right) \ ,
\]
viz.
\begin{knitrout}
\definecolor{shadecolor}{rgb}{0.969, 0.969, 0.969}\color{fgcolor}\begin{kframe}
\begin{alltt}
\hlkwd{round}\hlstd{(}\hlkwc{x} \hlstd{= (LambdaCor} \hlopt{-} \hlkwd{diag}\hlstd{(}\hlkwc{x} \hlstd{= evAnaCor}\hlopt{$}\hlstd{values)),} \hlkwc{digits} \hlstd{=} \hlnum{4}\hlstd{)}
\end{alltt}
\begin{verbatim}
##      [,1] [,2] [,3]
## [1,]    0    0    0
## [2,]    0    0    0
## [3,]    0    0    0
\end{verbatim}
\end{kframe}
\end{knitrout}

\medskip
\noindent
The \textbf{inverse}, $\boldsymbol{\Lambda}^{-1}$, of the 
\textbf{diagonal eigenvalue matrix}~$\boldsymbol{\Lambda}$ is computed by
\begin{knitrout}
\definecolor{shadecolor}{rgb}{0.969, 0.969, 0.969}\color{fgcolor}\begin{kframe}
\begin{alltt}
\hlstd{LambdaCorInv} \hlkwb{<-}
  \hlkwd{diag}\hlstd{(}\hlkwc{x} \hlstd{= (}\hlnum{1} \hlopt{/} \hlstd{evAnaCor}\hlopt{$}\hlstd{values))}
\hlstd{LambdaCorInv}
\end{alltt}
\begin{verbatim}
##           [,1]     [,2]     [,3]
## [1,] 0.5752941 0.000000 0.000000
## [2,] 0.0000000 1.131464 0.000000
## [3,] 0.0000000 0.000000 2.645865
\end{verbatim}
\end{kframe}
\end{knitrout}

\medskip
\noindent
The three \textbf{eigenvalues} of the \textbf{sample correlation
matrix}~$\boldsymbol{R}$ amount to the \textbf{variances} of the data in
$\boldsymbol{Z}$ along the three directions in Euclidian space $\mathbb{R}^{3}$
defined by the \textbf{orthonormal eigenbasis} of the \textbf{sample correlation
matrix} ~$\boldsymbol{R}$. The following consideration makes this fact explicit.

\subsection{Visualisation of data in $\boldsymbol{Z}_{\mathrm{rot}}$ via 3D
scatter plot}
\lb{subsec:Zrot3Dscatplot}
\textbf{Transformation} of the data in $\boldsymbol{Z}$ to the
\textbf{orthonormal eigenbasis} of the
\textbf{sample correlation matrix}~$\boldsymbol{R}$ yields
\[
\boldsymbol{Z}_{\mathrm{rot}} = \boldsymbol{Z}\boldsymbol{V} \ ,
\]
viz.
\begin{knitrout}
\definecolor{shadecolor}{rgb}{0.969, 0.969, 0.969}\color{fgcolor}\begin{kframe}
\begin{alltt}
\hlstd{Zrot} \hlkwb{<-}
  \hlstd{Z} \hlopt{%*%} \hlstd{rotMatCor} \hlopt{%>%}
  \hlstd{magrittr}\hlopt{::}\hlkwd{set_colnames}\hlstd{(}\hlkwc{x} \hlstd{= .,} \hlkwc{value} \hlstd{=} \hlkwd{c}\hlstd{(}\hlstr{"PC1"}\hlstd{,} \hlstr{"PC2"}\hlstd{,} \hlstr{"PC3"}\hlstd{))}
\end{alltt}
\end{kframe}
\end{knitrout}

\noindent
Visualisation of the resultant data in $\boldsymbol{Z}_{\mathrm{rot}}$ by
means of a \textbf{3D scatter plot} gives
\begin{knitrout}
\definecolor{shadecolor}{rgb}{0.969, 0.969, 0.969}\color{fgcolor}\begin{kframe}
\begin{alltt}
\hlstd{fig3} \hlkwb{<-}
  \hlstd{plotly}\hlopt{::}\hlkwd{plot_ly}\hlstd{(}
    \hlkwc{data} \hlstd{= tibble}\hlopt{::}\hlkwd{as_tibble}\hlstd{(}\hlkwc{x} \hlstd{= Zrot),}
    \hlkwc{type} \hlstd{=} \hlstr{"scatter3d"}\hlstd{,}
    \hlkwc{x} \hlstd{= Zrot[,} \hlnum{1}\hlstd{],}
    \hlkwc{y} \hlstd{= Zrot[,} \hlnum{2}\hlstd{],}
    \hlkwc{z} \hlstd{= Zrot[,} \hlnum{3}\hlstd{],}
    \hlkwc{mode} \hlstd{=} \hlstr{"markers"}\hlstd{,}
    \hlkwc{size} \hlstd{=} \hlnum{1}
  \hlstd{)} \hlopt{%>%}
  \hlstd{plotly}\hlopt{::}\hlkwd{layout}\hlstd{(}\hlkwc{title} \hlstd{=} \hlkwd{paste0}\hlstd{(}\hlstr{"Standardised data wrt. "}\hlstd{,}
                                \hlstr{"orthonormal eigenbasis of R"}\hlstd{),}
                 \hlkwc{scene} \hlstd{=} \hlkwd{list}\hlstd{(}
                   \hlkwc{xaxis} \hlstd{=} \hlkwd{list}\hlstd{(}\hlkwc{title} \hlstd{=} \hlstr{"Zrot1 [1]"}\hlstd{),}
                   \hlkwc{yaxis} \hlstd{=} \hlkwd{list}\hlstd{(}\hlkwc{title} \hlstd{=} \hlstr{"Zrot2 [1]"}\hlstd{),}
                   \hlkwc{zaxis} \hlstd{=} \hlkwd{list}\hlstd{(}\hlkwc{title} \hlstd{=} \hlstr{"Zrot3 [1]"}\hlstd{)}
                 \hlstd{))}
\hlstd{fig3}
\end{alltt}
\end{kframe}
\end{knitrout}

\medskip
{\centering \includegraphics[width=14cm]{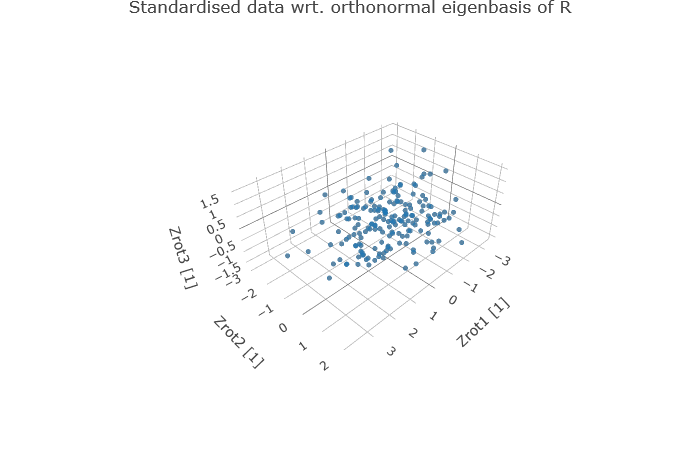}}

\medskip
\noindent
Subsequent computation of the \textbf{variances} of the data in
$\boldsymbol{Z}_{\mathrm{rot}}$ obtains
\begin{knitrout}
\definecolor{shadecolor}{rgb}{0.969, 0.969, 0.969}\color{fgcolor}\begin{kframe}
\begin{alltt}
\hlkwd{apply}\hlstd{(}\hlkwc{X} \hlstd{= Zrot,} \hlkwc{MARGIN} \hlstd{=} \hlnum{2}\hlstd{,} \hlkwc{FUN} \hlstd{= var)}
\end{alltt}
\begin{verbatim}
##       PC1       PC2       PC3 
## 1.7382412 0.8838105 0.3779482
\end{verbatim}
\end{kframe}
\end{knitrout}

\noindent
i.e., the claimed result. Note that the data in the columns of
$\boldsymbol{Z}_{\mathrm{rot}}$ is pairwise \textit{uncorrelated}
\begin{knitrout}
\definecolor{shadecolor}{rgb}{0.969, 0.969, 0.969}\color{fgcolor}\begin{kframe}
\begin{alltt}
\hlkwd{round}\hlstd{(}\hlkwc{x} \hlstd{=} \hlkwd{cor}\hlstd{(}\hlkwc{x} \hlstd{= Zrot),} \hlkwc{digits} \hlstd{=} \hlnum{4}\hlstd{)}
\end{alltt}
\begin{verbatim}
##     PC1 PC2 PC3
## PC1   1   0   0
## PC2   0   1   0
## PC3   0   0   1
\end{verbatim}
\end{kframe}
\end{knitrout}
%

\subsection{Visualisation of data in $\boldsymbol{Z}_{\mathrm{rot}}$ via scatter
plot matrix}
\lb{subsec:Zrotscatplotmat}
%
\begin{knitrout}
\definecolor{shadecolor}{rgb}{0.969, 0.969, 0.969}\color{fgcolor}\begin{kframe}
\begin{alltt}
\hlstd{GGally}\hlopt{::}\hlkwd{ggpairs}\hlstd{(}\hlkwc{data} \hlstd{= tibble}\hlopt{::}\hlkwd{as_tibble}\hlstd{(}\hlkwc{x} \hlstd{= Zrot))} \hlopt{+}
  \hlkwd{theme_bw}\hlstd{()}
\end{alltt}
\end{kframe}

{\centering \includegraphics[width=\maxwidth]{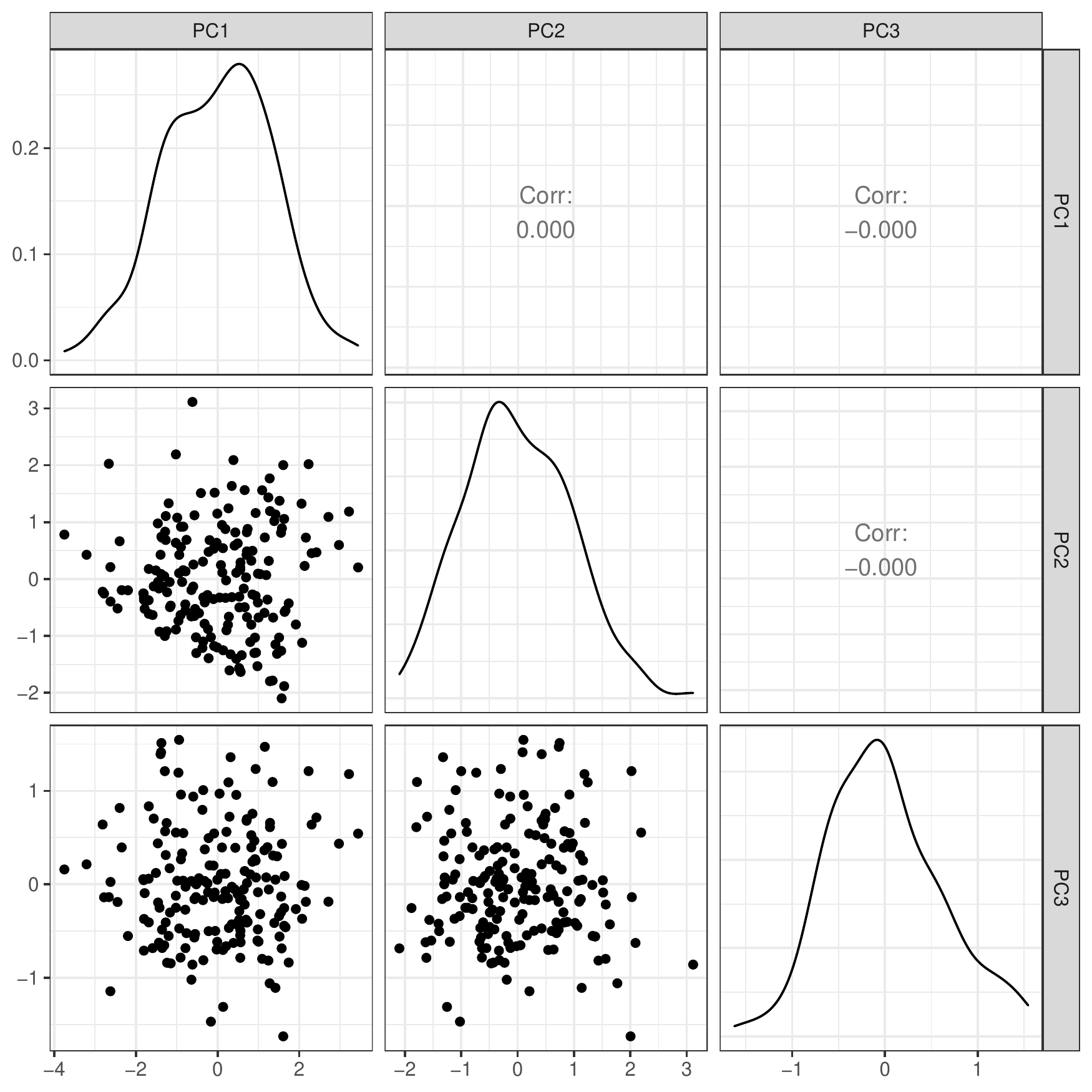} 

}

\end{knitrout}
%

\subsection{Visualisation of data in $\boldsymbol{Z}_{\mathrm{rot}}$ via
parallel box plots}
\lb{subsec:Zrotboxplots}
%
\begin{knitrout}
\definecolor{shadecolor}{rgb}{0.969, 0.969, 0.969}\color{fgcolor}\begin{kframe}
\begin{alltt}
\hlstd{Zrot} \hlopt{%>%}
  \hlstd{tibble}\hlopt{::}\hlkwd{as_tibble}\hlstd{(}\hlkwc{x} \hlstd{= .)} \hlopt{%>%}
  \hlstd{tidyr}\hlopt{::}\hlkwd{pivot_longer}\hlstd{(}
    \hlkwc{data} \hlstd{= .,}
    \hlkwc{cols} \hlstd{=} \hlkwd{c}\hlstd{(`PC1`, `PC2`, `PC3`),}
    \hlkwc{names_to} \hlstd{=} \hlstr{"dimension"}\hlstd{,}
    \hlkwc{values_to} \hlstd{=} \hlstr{"zscorerot"}
  \hlstd{)} \hlopt{%>%}
  \hlstd{dplyr}\hlopt{::}\hlkwd{mutate}\hlstd{(}
    \hlkwc{dimension} \hlstd{= forcats}\hlopt{::}\hlkwd{fct_relevel}\hlstd{(}
      \hlkwc{.f} \hlstd{= dimension,}
      \hlkwd{c}\hlstd{(}\hlstr{"PC1"}\hlstd{,} \hlstr{"PC2"}\hlstd{,} \hlstr{"PC3"}\hlstd{))}
    \hlstd{)} \hlopt{%>%}
  \hlkwd{ggplot}\hlstd{(}\hlkwc{data} \hlstd{= .,}
         \hlkwc{mapping} \hlstd{=} \hlkwd{aes}\hlstd{(}\hlkwc{x} \hlstd{= dimension,} \hlkwc{y} \hlstd{= zscorerot))} \hlopt{+}
  \hlkwd{geom_boxplot}\hlstd{()} \hlopt{+}
  \hlkwd{xlab}\hlstd{(}\hlkwc{label} \hlstd{=} \hlstr{"dimension"}\hlstd{)} \hlopt{+}
  \hlkwd{ylab}\hlstd{(}\hlkwc{label} \hlstd{=} \hlstr{"rotated zscore [1]"}\hlstd{)} \hlopt{+}
  \hlkwd{theme_bw}\hlstd{()}
\end{alltt}
\end{kframe}

{\centering \includegraphics[width=\maxwidth]{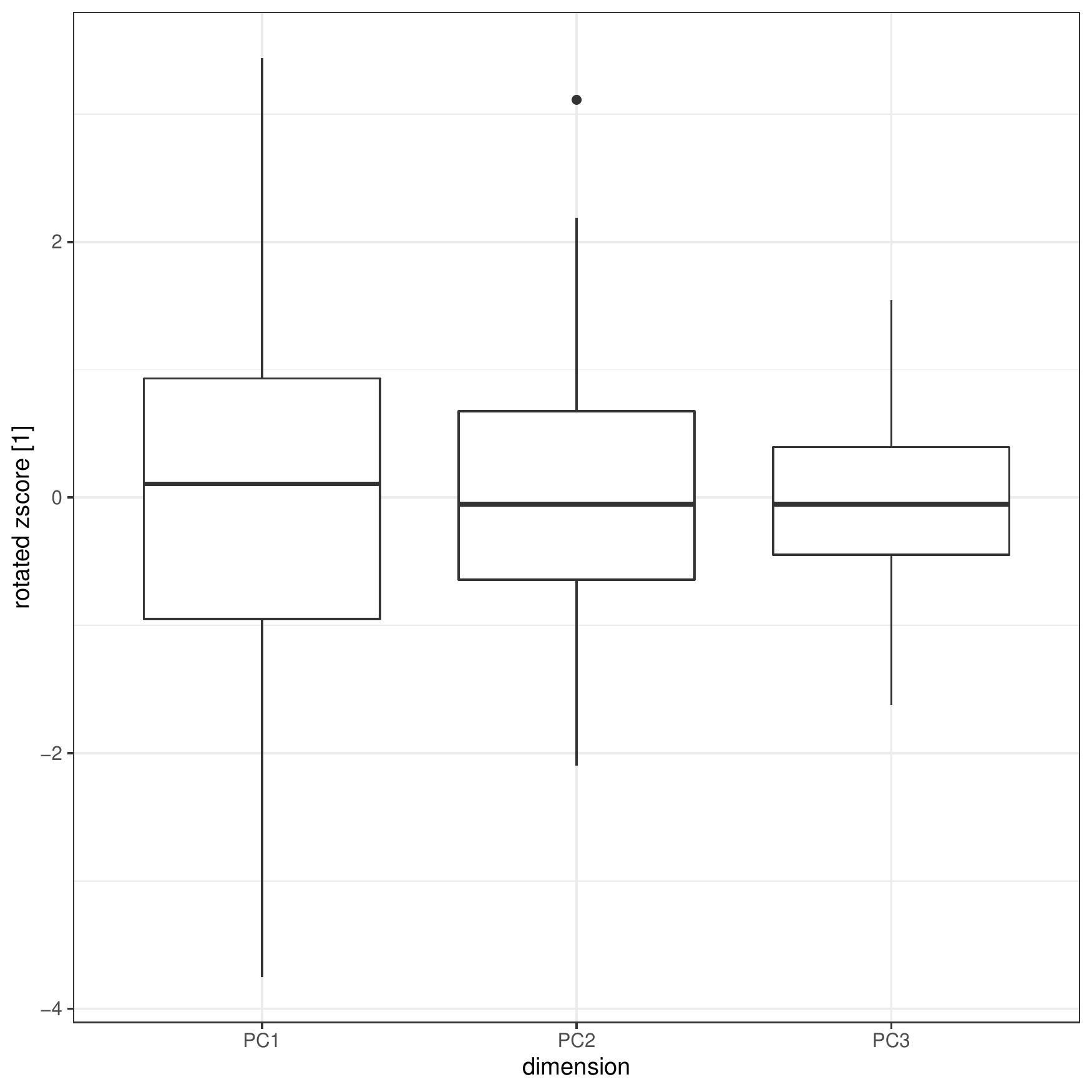} 

}

\end{knitrout}

\medskip
\noindent
Note that in the \textbf{orthonormal eigenbasis} of the \textbf{sample
correlation matrix}~$\boldsymbol{R}$ the \textbf{heterogeneity of variances}
between the univariate data for each of the three variables in the rotated
\textbf{data matrix}~$\boldsymbol{Z}_{\mathrm{rot}}$ is more pronounced
than the corresponding one in the \textbf{original orthonormal basis} between
the univariate data for each of the three variables in the
\textbf{data matrix}~$\boldsymbol{Z}$. This is a clear manifestation of the
existing \textbf{correlations} between the latter three variables.

\subsection{Descriptive statistics for data in $\boldsymbol{Z}_{\mathrm{rot}}$}
\lb{subsec:Zrotdescrstats}
Descriptive statistical measures are computed that characterise the observed
univariate distributions for the data in $\boldsymbol{Z}_{\mathrm{rot}}$
(``rotated $z$-scores''):
\begin{enumerate}

\item \textbf{Mean} and \textbf{standard deviation}:
\begin{knitrout}
\definecolor{shadecolor}{rgb}{0.969, 0.969, 0.969}\color{fgcolor}\begin{kframe}
\begin{alltt}
\hlkwd{apply}\hlstd{(}\hlkwc{X} \hlstd{= Zrot,} \hlkwc{MARGIN} \hlstd{=} \hlnum{2}\hlstd{,} \hlkwc{FUN} \hlstd{= mean)} \hlopt{%>%}
  \hlkwd{round}\hlstd{(}\hlkwc{x} \hlstd{= .,} \hlkwc{digits} \hlstd{=} \hlnum{4}\hlstd{)}
\end{alltt}
\begin{verbatim}
## PC1 PC2 PC3 
##   0   0   0
\end{verbatim}
\begin{alltt}
\hlkwd{apply}\hlstd{(}\hlkwc{X} \hlstd{= Zrot,} \hlkwc{MARGIN} \hlstd{=} \hlnum{2}\hlstd{,} \hlkwc{FUN} \hlstd{= sd)}
\end{alltt}
\begin{verbatim}
##      PC1      PC2      PC3 
## 1.318424 0.940112 0.614775
\end{verbatim}
\end{kframe}
\end{knitrout}

\item \textbf{Standardised skewness} and \textbf{standardised excess kurtosis};
cf. Joanes and Gill (1998)~\ct{joagil1998} and van Elst (2019)~\ct{hve2019}:
\begin{knitrout}
\definecolor{shadecolor}{rgb}{0.969, 0.969, 0.969}\color{fgcolor}\begin{kframe}
\begin{alltt}
\hlkwd{standSkewness}\hlstd{(Zrot)}
\end{alltt}
\begin{verbatim}
##        PC1        PC2        PC3 
## -0.6580145  1.5543143  1.9033877
\end{verbatim}
\begin{alltt}
\hlkwd{standKurtosis}\hlstd{(Zrot)}
\end{alltt}
\begin{verbatim}
##        PC1        PC2        PC3 
## -0.5119328 -0.3687139 -0.0712973
\end{verbatim}
\end{kframe}
\end{knitrout}

\end{enumerate}
%

\section[Principal component loadings matrix]{Principal component loadings
matrix $\boldsymbol{A}$}
\lb{sec:loadmat}
The \textbf{principal component loadings matrix}~$\boldsymbol{A}$, defined
in terms of the orthogonal \textbf{rotation matrix}~$\boldsymbol{V}$ and the
\textbf{diagonal eigenvalue matrix}~$\boldsymbol{\Lambda}$ by
\[
\boldsymbol{A} := \boldsymbol{V}\boldsymbol{\Lambda}^{1/2} \ ,
\]
provides the answer to the question: how strongly are the (presently) three
\textbf{original variables} height, mass and age correlated with the identified
three \textbf{principal components} of the \textbf{trivariate data set}
considered?\footnote{Generally: how strongly are the $m$~original variables
of a given metrically scaled multivariate data set correlated with the
identified $m$~principal components of this data set?}
\begin{knitrout}
\definecolor{shadecolor}{rgb}{0.969, 0.969, 0.969}\color{fgcolor}\begin{kframe}
\begin{alltt}
\hlstd{AmatCor} \hlkwb{<-}
  \hlstd{rotMatCor} \hlopt{%*%} \hlkwd{abs}\hlstd{(LambdaCor)} \hlopt{^} \hlstd{(}\hlnum{1} \hlopt{/} \hlnum{2}\hlstd{)} \hlopt{%>%}
  \hlstd{magrittr}\hlopt{::}\hlkwd{set_rownames}\hlstd{(}
    \hlkwc{x} \hlstd{= .,}
    \hlkwc{value} \hlstd{=} \hlkwd{c}\hlstd{(}\hlstr{"height"}\hlstd{,} \hlstr{"mass"}\hlstd{,} \hlstr{"age"}\hlstd{))} \hlopt{%>%}
  \hlstd{magrittr}\hlopt{::}\hlkwd{set_colnames}\hlstd{(}
    \hlkwc{x} \hlstd{= .,}
    \hlkwc{value} \hlstd{=} \hlkwd{c}\hlstd{(}\hlstr{"PC1"}\hlstd{,} \hlstr{"PC2"}\hlstd{,} \hlstr{"PC3"}\hlstd{))}
\hlstd{AmatCor}
\end{alltt}
\begin{verbatim}
##               PC1       PC2         PC3
## height -0.8575507 0.2852016 -0.42809679
## mass   -0.8735813 0.2083200  0.43983924
## age     0.4895956 0.8712482  0.03496895
\end{verbatim}
\end{kframe}
\end{knitrout}

\noindent
The fact that the \textbf{principal component loadings matrix}~$\boldsymbol{A}$
may indeed be interpreted as a (formal) correlation matrix will become apparent 
later on.

\medskip
\noindent
The \textbf{principal component loadings matrix}~$\boldsymbol{A}$ satisfies
two \textbf{consistency checks}:
\begin{enumerate}

\item The \textbf{sample correlation matrix}~$\boldsymbol{R}$ can be factorised
by means of the \textbf{principal component loadings matrix}~$\boldsymbol{A}$ as
\[
\boldsymbol{0} = \boldsymbol{R}-\boldsymbol{A}\boldsymbol{A}^{\top} \ ,
\]
viz.
\begin{knitrout}
\definecolor{shadecolor}{rgb}{0.969, 0.969, 0.969}\color{fgcolor}\begin{kframe}
\begin{alltt}
\hlkwd{round}\hlstd{(}\hlkwc{x} \hlstd{= (Rmat} \hlopt{-} \hlstd{AmatCor} \hlopt{%*%} \hlkwd{t}\hlstd{(AmatCor)),} \hlkwc{digits} \hlstd{=} \hlnum{4}\hlstd{)}
\end{alltt}
\begin{verbatim}
##                height_std [1] mass_std [1] age_std [1]
## height_std [1]              0            0           0
## mass_std [1]                0            0           0
## age_std [1]                 0            0           0
\end{verbatim}
\end{kframe}
\end{knitrout}

\item The \textbf{diagonal eigenvalue matrix}~$\boldsymbol{\Lambda}$ can be
factorised by means of the \textbf{principal component loadings
matrix}~$\boldsymbol{A}$ as
\[
\boldsymbol{0} = \boldsymbol{\Lambda}-\boldsymbol{A}^{\top}\boldsymbol{A} \ ,
\]
viz.
\begin{knitrout}
\definecolor{shadecolor}{rgb}{0.969, 0.969, 0.969}\color{fgcolor}\begin{kframe}
\begin{alltt}
\hlkwd{round}\hlstd{(}\hlkwc{x} \hlstd{= (LambdaCor} \hlopt{-} \hlkwd{t}\hlstd{(AmatCor)} \hlopt{%*%} \hlstd{AmatCor),} \hlkwc{digits} \hlstd{=} \hlnum{4}\hlstd{)}
\end{alltt}
\begin{verbatim}
##     PC1 PC2 PC3
## PC1   0   0   0
## PC2   0   0   0
## PC3   0   0   0
\end{verbatim}
\end{kframe}
\end{knitrout}

\end{enumerate}
%

\section[Standardised data matrix in orthonormal eigenbasis]{Standardised
data set in orthonormal eigenbasis of sample correlation matrix
(matrix $\boldsymbol{F}$)}
\lb{sec:fmat}
Finally, a \textbf{transformation} needs to be performed of the standardised 
\textbf{trivariate data set} in $\boldsymbol{Z}$ to the \textbf{orthonormal 
eigenbasis} of the \textbf{sample correlation matrix}~$\boldsymbol{R}$, while
respecting the conventional requirement that the resultant data shall likewise
be standardised. This is realised in terms of a volume preserving 
\textbf{rotation} of the original reference frame with the \textbf{rotation 
matrix}~$\boldsymbol{V}$,\footnote{The effect of a pure transformation of the
data in $\boldsymbol{Z}$ with the rotation matrix~$\boldsymbol{V}$ was described
and visualised before in Sec.~\ref{sec:rotmat}.} followed by a volume changing
but directions preserving \textbf{rescaling} of the axes of the rotated
reference frame with the (square root of the)
\textbf{inverse}~$\boldsymbol{\Lambda}^{-1}$ of the \textbf{diagonal eigenvalue 
matrix}~$\boldsymbol{\Lambda}$. These two \textbf{transformations} performed in
combination define the matrix~$\boldsymbol{F}$,\footnote{Equivalently, the 
matrix~$\boldsymbol{F}$ can also be computed from the trivariate data set in 
$\boldsymbol{Z}$ by employing the principal component loadings
matrix~$\boldsymbol{A}$ and the inverse~$\boldsymbol{\Lambda}^{-1}$
as $\boldsymbol{F} = \boldsymbol{Z}\boldsymbol{A}\boldsymbol{\Lambda}^{-1}$.}
\[
\boldsymbol{F} := \boldsymbol{Z}\boldsymbol{V}\boldsymbol{\Lambda}^{-1/2} \ ,
\]
viz.
\begin{knitrout}
\definecolor{shadecolor}{rgb}{0.969, 0.969, 0.969}\color{fgcolor}\begin{kframe}
\begin{alltt}
\hlstd{FmatCor} \hlkwb{<-}
  \hlstd{Z} \hlopt{%*%} \hlstd{rotMatCor} \hlopt{%*%} \hlstd{LambdaCorInv} \hlopt{^} \hlstd{(}\hlnum{1} \hlopt{/} \hlnum{2}\hlstd{)} \hlopt{%>%}
  \hlstd{magrittr}\hlopt{::}\hlkwd{set_colnames}\hlstd{(}
    \hlkwc{x} \hlstd{= .,}
    \hlkwc{value} \hlstd{=} \hlkwd{c}\hlstd{(}\hlstr{"PC1_std [1]"}\hlstd{,} \hlstr{"PC2_std [1]"}\hlstd{,} \hlstr{"PC3_std [1]"}\hlstd{))}
\hlkwd{dim}\hlstd{(}\hlkwc{x} \hlstd{= FmatCor)}
\end{alltt}
\begin{verbatim}
## [1] 187   3
\end{verbatim}
\begin{alltt}
\hlkwd{head}\hlstd{(}\hlkwc{x} \hlstd{= FmatCor)}
\end{alltt}
\begin{verbatim}
##      PC1_std [1] PC2_std [1] PC3_std [1]
## [1,]   1.8362245   0.4986426   1.1623874
## [2,]   2.6100451   0.2165376   0.8829885
## [3,]   0.6264361   0.3416280   0.8556499
## [4,]   0.2097674  -0.7040217  -0.7566757
## [5,]   0.4219218  -1.7223008  -1.2765885
## [6,]  -1.1094795   1.0397559   0.7139017
\end{verbatim}
\end{kframe}
\end{knitrout}

\noindent
which contains standardised and, by construction, mutually
\textit{uncorrelated} so-called ``$f$-scores'' in its columns.

\subsection{Consistency checks for matrix $\boldsymbol{F}$}
The following \textbf{consistency checks} need to be satisfied by the
matrix~$\boldsymbol{F}$:
\begin{enumerate}

\item The ``$f$-scores'' are standardised and mutually uncorrelated:
\[
{\displaystyle \boldsymbol{0} = \boldsymbol{1} - \frac{1}{n-1}\,
\boldsymbol{F}^{\top}\boldsymbol{F}} \ ,
\]
viz.
\begin{knitrout}
\definecolor{shadecolor}{rgb}{0.969, 0.969, 0.969}\color{fgcolor}\begin{kframe}
\begin{alltt}
\hlstd{proxy} \hlkwb{<-}
  \hlstd{(}\hlnum{1} \hlopt{/} \hlstd{(}\hlkwd{nrow}\hlstd{(FmatCor)} \hlopt{-} \hlnum{1}\hlstd{))} \hlopt{*} \hlkwd{t}\hlstd{(FmatCor)} \hlopt{%*%} \hlstd{FmatCor}
\hlkwd{round}\hlstd{(}\hlkwc{x} \hlstd{= (}\hlkwd{diag}\hlstd{(}\hlkwd{rep}\hlstd{(}\hlkwc{x} \hlstd{=} \hlnum{1}\hlstd{,} \hlkwc{times} \hlstd{=} \hlkwd{nrow}\hlstd{(proxy)))} \hlopt{-} \hlstd{proxy),}
      \hlkwc{digits} \hlstd{=} \hlnum{4}\hlstd{)}
\end{alltt}
\begin{verbatim}
##             PC1_std [1] PC2_std [1] PC3_std [1]
## PC1_std [1]           0           0           0
## PC2_std [1]           0           0           0
## PC3_std [1]           0           0           0
\end{verbatim}
\end{kframe}
\end{knitrout}

\item The elements of the \textbf{principal component loadings
matrix}~$\boldsymbol{A}$ represent, as algebraic projections of standardised
and uncorrelated ``$z$-scores'' onto standardised and uncorrelated
``$f$-scores'', bivariate correlations between the (presently three)
\textbf{original variables} and the (presently three) \textbf{principal
components} of the \textbf{multivariate data set} considered; cf. the remarks
at the beginning of Sec.~\ref{sec:loadmat}:
\[
{\displaystyle \boldsymbol{0} = \boldsymbol{A} - \frac{1}{n-1}\,
\boldsymbol{Z}^{\top}\boldsymbol{F}} \ ,
\]
viz.
\begin{knitrout}
\definecolor{shadecolor}{rgb}{0.969, 0.969, 0.969}\color{fgcolor}\begin{kframe}
\begin{alltt}
\hlkwd{round}\hlstd{(}
  \hlkwc{x} \hlstd{= (AmatCor} \hlopt{-} \hlstd{(}\hlnum{1} \hlopt{/} \hlstd{(}\hlkwd{nrow}\hlstd{(Z)} \hlopt{-} \hlnum{1}\hlstd{))} \hlopt{*} \hlkwd{t}\hlstd{(Z)} \hlopt{%*%} \hlstd{FmatCor),}
  \hlkwc{digits} \hlstd{=} \hlnum{4}
  \hlstd{)}
\end{alltt}
\begin{verbatim}
##        PC1 PC2 PC3
## height   0   0   0
## mass     0   0   0
## age      0   0   0
\end{verbatim}
\end{kframe}
\end{knitrout}

\item The ``$z$-scores'' may be perceived as linear combinations of the
``$f$-scores'':
\[{\displaystyle \boldsymbol{0} = \boldsymbol{Z} - \boldsymbol{F}
\boldsymbol{A}^{\top}} \ ,
\]
viz.
\begin{knitrout}
\definecolor{shadecolor}{rgb}{0.969, 0.969, 0.969}\color{fgcolor}\begin{kframe}
\begin{alltt}
\hlkwd{head}\hlstd{(}\hlkwc{x} \hlstd{=} \hlkwd{round}\hlstd{(}\hlkwc{x} \hlstd{= (Z} \hlopt{-} \hlstd{FmatCor} \hlopt{%*%} \hlkwd{t}\hlstd{(AmatCor)),} \hlkwc{digits} \hlstd{=} \hlnum{4}\hlstd{))}
\end{alltt}
\begin{verbatim}
##      height_std [1] mass_std [1] age_std [1]
## [1,]              0            0           0
## [2,]              0            0           0
## [3,]              0            0           0
## [4,]              0            0           0
## [5,]              0            0           0
## [6,]              0            0           0
\end{verbatim}
\begin{alltt}
\hlkwd{tail}\hlstd{(}\hlkwc{x} \hlstd{=} \hlkwd{round}\hlstd{(}\hlkwc{x} \hlstd{= (Z} \hlopt{-} \hlstd{FmatCor} \hlopt{%*%} \hlkwd{t}\hlstd{(AmatCor)),} \hlkwc{digits} \hlstd{=} \hlnum{4}\hlstd{))}
\end{alltt}
\begin{verbatim}
##        height_std [1] mass_std [1] age_std [1]
## [182,]              0            0           0
## [183,]              0            0           0
## [184,]              0            0           0
## [185,]              0            0           0
## [186,]              0            0           0
## [187,]              0            0           0
\end{verbatim}
\end{kframe}
\end{knitrout}

\end{enumerate}
%

\subsection{Visualisation of data in $\boldsymbol{F}$ via 3D scatter plot}
%
\begin{knitrout}
\definecolor{shadecolor}{rgb}{0.969, 0.969, 0.969}\color{fgcolor}\begin{kframe}
\begin{alltt}
\hlstd{fig4} \hlkwb{<-}
  \hlstd{plotly}\hlopt{::}\hlkwd{plot_ly}\hlstd{(}
    \hlkwc{data} \hlstd{= tibble}\hlopt{::}\hlkwd{as_tibble}\hlstd{(}\hlkwc{x} \hlstd{= FmatCor),}
    \hlkwc{type} \hlstd{=} \hlstr{"scatter3d"}\hlstd{,}
    \hlkwc{x} \hlstd{= FmatCor[,} \hlnum{1}\hlstd{],}
    \hlkwc{y} \hlstd{= FmatCor[,} \hlnum{2}\hlstd{],}
    \hlkwc{z} \hlstd{= FmatCor[,} \hlnum{3}\hlstd{],}
    \hlkwc{mode} \hlstd{=} \hlstr{"markers"}\hlstd{,}
    \hlkwc{size} \hlstd{=} \hlnum{1}
  \hlstd{)} \hlopt{%>%}
  \hlstd{plotly}\hlopt{::}\hlkwd{layout}\hlstd{(}\hlkwc{title} \hlstd{=} \hlkwd{paste0}\hlstd{(}\hlstr{"Standardised data in F "}\hlstd{,}
                                \hlstr{"(unit scale of measurement)"}\hlstd{),}
                 \hlkwc{scene} \hlstd{=} \hlkwd{list}\hlstd{(}
                   \hlkwc{xaxis} \hlstd{=} \hlkwd{list}\hlstd{(}\hlkwc{title} \hlstd{=} \hlstr{"PC1 std [1]"}\hlstd{),}
                   \hlkwc{yaxis} \hlstd{=} \hlkwd{list}\hlstd{(}\hlkwc{title} \hlstd{=} \hlstr{"PC2 std [1]"}\hlstd{),}
                   \hlkwc{zaxis} \hlstd{=} \hlkwd{list}\hlstd{(}\hlkwc{title} \hlstd{=} \hlstr{"PC3 std [1]"}\hlstd{)}
                 \hlstd{))}
\hlstd{fig4}
\end{alltt}
\end{kframe}
\end{knitrout}

\medskip
{\centering \includegraphics[width=14cm]{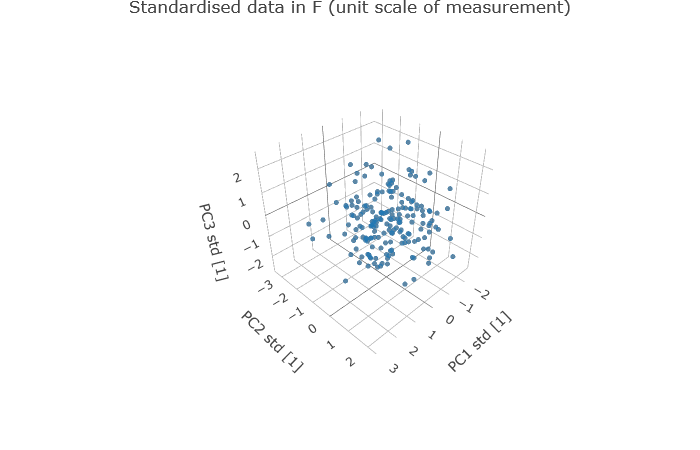}}
%

\subsection{Visualisation of data in $\boldsymbol{F}$ via scatter plot matrix}
%
\begin{knitrout}
\definecolor{shadecolor}{rgb}{0.969, 0.969, 0.969}\color{fgcolor}\begin{kframe}
\begin{alltt}
\hlstd{GGally}\hlopt{::}\hlkwd{ggpairs}\hlstd{(}\hlkwc{data} \hlstd{= tibble}\hlopt{::}\hlkwd{as_tibble}\hlstd{(}\hlkwc{x} \hlstd{= FmatCor))} \hlopt{+}
  \hlkwd{theme_bw}\hlstd{()}
\end{alltt}
\end{kframe}

{\centering \includegraphics[width=\maxwidth]{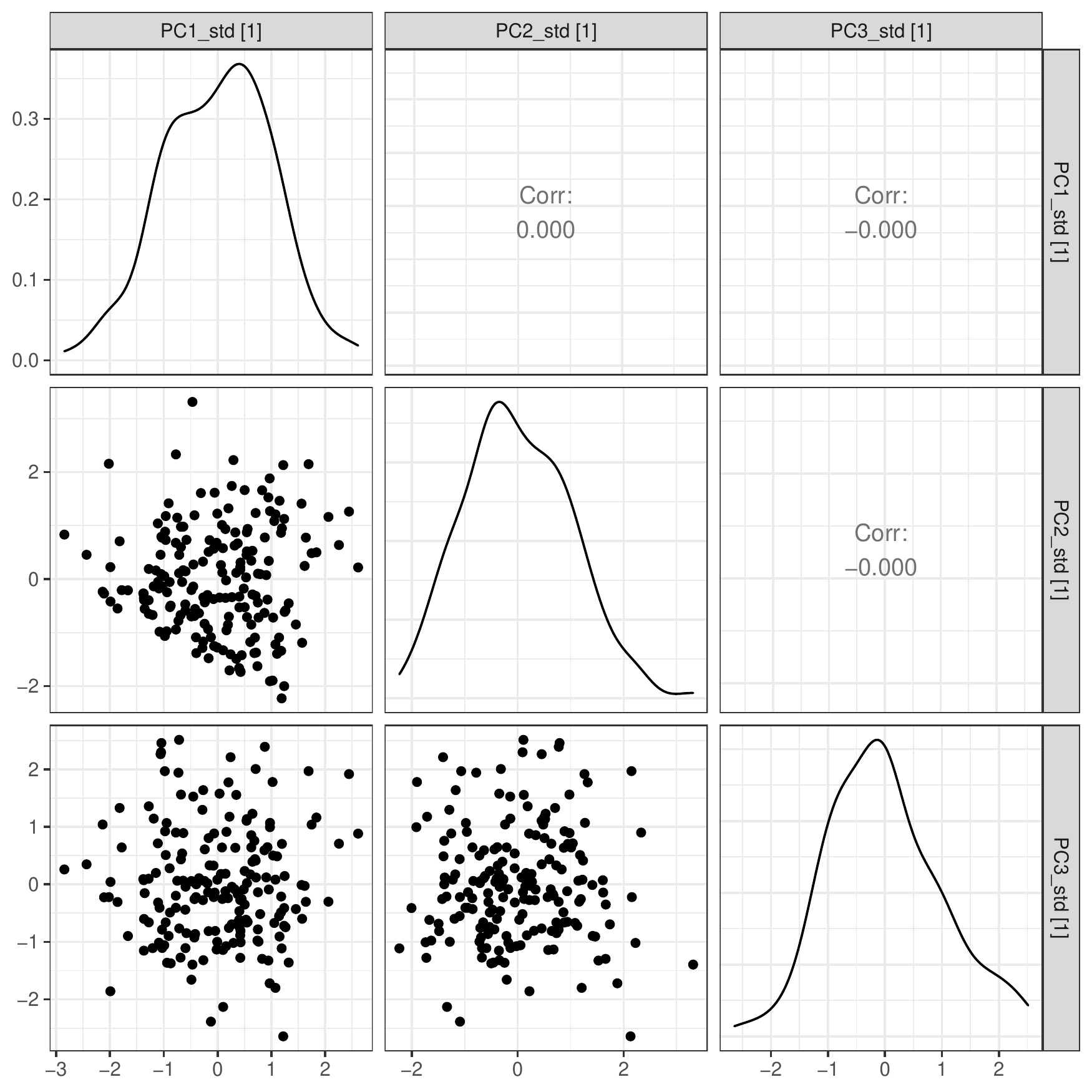} 

}

\end{knitrout}

\medskip
\noindent
The \textbf{scatter plot matrix} shows nicely that the ``$f$-scores''
associated with each of the \textbf{principal components} are mutually
\textit{uncorrelated}.

\subsection{Visualisation of data in $\boldsymbol{F}$ via
parallel box plots}
\lb{subsec:Fboxplots}
%
\begin{knitrout}
\definecolor{shadecolor}{rgb}{0.969, 0.969, 0.969}\color{fgcolor}\begin{kframe}
\begin{alltt}
\hlstd{FmatCor} \hlopt{%>%}
  \hlstd{tibble}\hlopt{::}\hlkwd{as_tibble}\hlstd{(}\hlkwc{x} \hlstd{= .)} \hlopt{%>%}
  \hlstd{tidyr}\hlopt{::}\hlkwd{pivot_longer}\hlstd{(}
    \hlkwc{data} \hlstd{= .,}
    \hlkwc{cols} \hlstd{=} \hlkwd{c}\hlstd{(`PC1_std [1]`, `PC2_std [1]`, `PC3_std [1]`),}
    \hlkwc{names_to} \hlstd{=} \hlstr{"dimension"}\hlstd{,}
    \hlkwc{values_to} \hlstd{=} \hlstr{"fscore"}
  \hlstd{)} \hlopt{%>%}
  \hlstd{dplyr}\hlopt{::}\hlkwd{mutate}\hlstd{(}
    \hlkwc{dimension} \hlstd{= forcats}\hlopt{::}\hlkwd{fct_relevel}\hlstd{(}
      \hlkwc{.f} \hlstd{= dimension,}
      \hlkwd{c}\hlstd{(}\hlstr{"PC1_std [1]"}\hlstd{,} \hlstr{"PC2_std [1]"}\hlstd{,} \hlstr{"PC3_std [1]"}\hlstd{))}
    \hlstd{)} \hlopt{%>%}
  \hlkwd{ggplot}\hlstd{(}\hlkwc{data} \hlstd{= .,}
         \hlkwc{mapping} \hlstd{=} \hlkwd{aes}\hlstd{(}\hlkwc{x} \hlstd{= dimension,} \hlkwc{y} \hlstd{= fscore))} \hlopt{+}
  \hlkwd{geom_boxplot}\hlstd{()} \hlopt{+}
  \hlkwd{xlab}\hlstd{(}\hlkwc{label} \hlstd{=} \hlstr{"dimension"}\hlstd{)} \hlopt{+}
  \hlkwd{ylab}\hlstd{(}\hlkwc{label} \hlstd{=} \hlstr{"fscore [1]"}\hlstd{)} \hlopt{+}
  \hlkwd{theme_bw}\hlstd{()}
\end{alltt}
\end{kframe}

{\centering \includegraphics[width=\maxwidth]{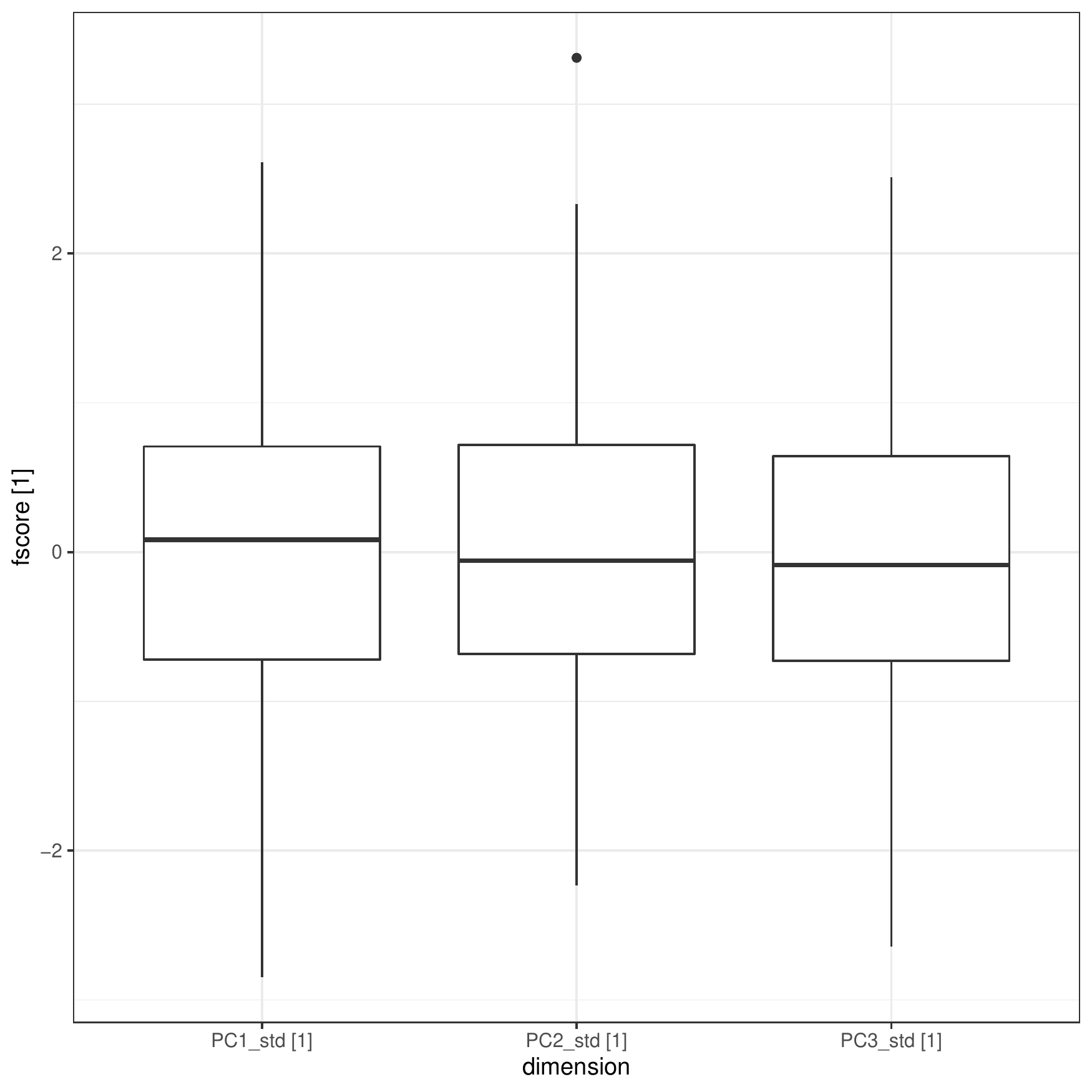} 

}

\end{knitrout}

\medskip
\noindent
\textbf{Standardisation} of the univariate data for each of the three variables
in the \textbf{orthonormal eigenbasis} of the \textbf{sample
correlation matrix}~$\boldsymbol{R}$ has neutralised the \textbf{heterogeneity
of variances} observed in Subsec.~\ref{subsec:Zrotboxplots}.

\section[Dimensional reduction]{Dimensional reduction: extraction of single
dominant principal component}
\lb{sec:dimred}
Secs.~\ref{sec:datastd} to~\ref{sec:fmat} featured in some detail the
linear-algebraic methodology on which the \textbf{principal component analysis}
of a metrically scaled \textbf{multivariate data set} is grounded. The
discussion now turns to describe the necessary steps involved in a
\textbf{dimensional reduction} of such a \textbf{data set}, given that for the
case considered the procedure proves conceptually and practically meaningful.

\medskip
\noindent
The \textbf{dimensional reduction} of the \textbf{multivariate data set} begins
in the \textbf{orthonormal eigenbasis} of the \textbf{sample correlation
matrix}~$\boldsymbol{R}$. In short, one only keeps those ``$f$-scores'' which
are associated with the \textbf{dominant principal components}; the remaining
ones are being discarded. The loss of information incurred in this way is
generally perceived as being compensated by an overall reduction of complexity
for the \textbf{multivariate data set} being analysed. The
remaining ``$f$-scores'' will be transformed back to the \textbf{original
orthonormal basis} and then expressed with respect to the \textbf{original
scales of measurement}. The \textbf{dimensionally reduced measured values} so 
obtained may form the starting point of further \textbf{statistical data
analysis}.\footnote{The performance of a dimensional reduction of a given
multivariate data set, when meaningful, does not really require computation of
the matrix~$\boldsymbol{F}$, respectively the ``$f$-scores''. For this purpose
it fully suffices to transform the ``$z$-scores'' from the original
orthonormal basis to the orthonormal eigenbasis of the sample correlation
matrix~$\boldsymbol{R}$ and then keep only those values (columns) associated
with the dominant principal components. Standardisation of values in the
orthonormal eigenbasis merely corresponds to a convenient convention.}

\subsection{Qualitative criterion for extraction}
The \R{} package $\texttt{psych}$ provides the function $\texttt{VSS.scree()}$
for generating \textbf{scree plots} according to Cattell (1966)~\ct{cat1966}.
\begin{knitrout}
\definecolor{shadecolor}{rgb}{0.969, 0.969, 0.969}\color{fgcolor}\begin{kframe}
\begin{alltt}
\hlstd{psych}\hlopt{::}\hlkwd{VSS.scree}\hlstd{(}\hlkwc{rx} \hlstd{= Z)}
\end{alltt}
\end{kframe}

{\centering \includegraphics[width=\maxwidth]{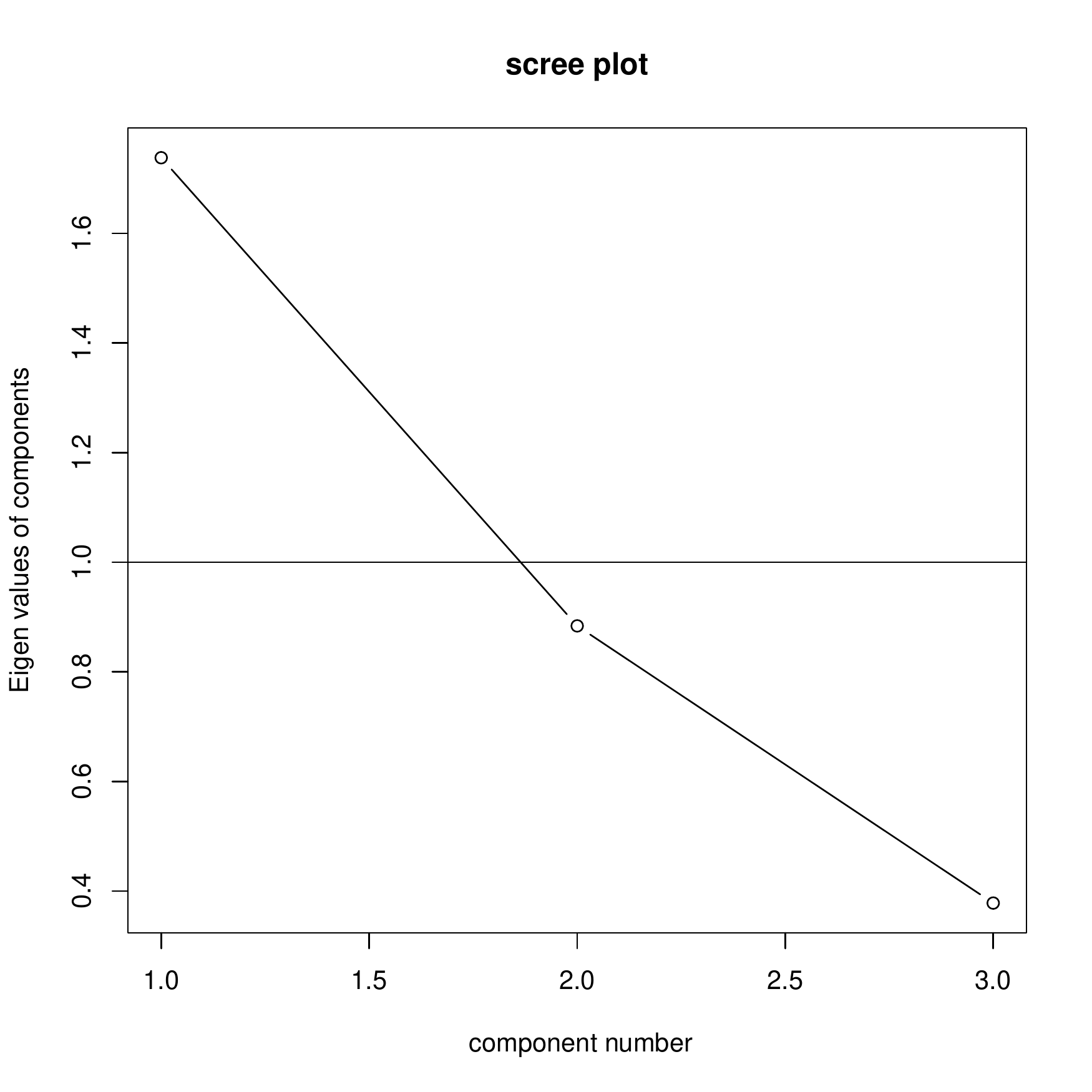} 

}

\end{knitrout}

\noindent
Based on numerous applications in practice, it is recommended to extract as many  
\textbf{principal components} fom a given \textbf{multivariate data set} as
correspond to the number of \textbf{eigenvalues} to be found \textit{left} of
the ``ellbow'' in a \textbf{scree plot}. In the present example this is just
one, and thus proves consistent with Kaiser's (1960)~\ct{kai1960}
\textbf{eigenvalue criterion}, which recommends \textbf{extraction} of all 
\textbf{principal components} that are associated with an
\textbf{eigenvalue} greater than $1$.

\medskip
\noindent
The procedure of a \textbf{dimensional reduction} of a given
\textbf{multivariate data set} is now exemplified for the \textbf{trivariate
data set} in $\boldsymbol{X}$. It is demonstrated on the basis of
\textbf{extraction} of the single \textbf{dominant principal component}
identified in Sec.~\ref{sec:eigen}.

\subsection{Dimensionally reduced matrixes}
The procedure of a \textbf{dimensional reduction} of a given multivariate data
set is reflected in the first place in the matrixes of the aforementioned
linear-algebraic methodology.
\begin{enumerate}

\item Dimensionally reduced rotation matrix~$\boldsymbol{V}_{\mathrm{red}}$
--- only those eigenvectors of a sample correlation matrix~$\boldsymbol{R}$ that
are associated with eigenvalues greater than $1$ will be employed in the
construction of a rotation matrix:
\begin{knitrout}
\definecolor{shadecolor}{rgb}{0.969, 0.969, 0.969}\color{fgcolor}\begin{kframe}
\begin{alltt}
\hlstd{rotMatCorRed} \hlkwb{<-}
  \hlstd{evAnaCor}\hlopt{$}\hlstd{vectors[,} \hlnum{1}\hlstd{]} \hlopt{%>%}
  \hlkwd{as.matrix}\hlstd{(}\hlkwc{x} \hlstd{= .)}
\hlstd{rotMatCorRed}
\end{alltt}
\begin{verbatim}
##            [,1]
## [1,] -0.6504363
## [2,] -0.6625952
## [3,]  0.3713492
\end{verbatim}
\end{kframe}
\end{knitrout}

\item Dimensionally reduced diagonal eigenvalue
matrix~$\boldsymbol{\Lambda}_{\mathrm{red}}$ --- generated from the sample
correlation matrix~$\boldsymbol{R}$ via transformation with the
dimensionally reduced rotation matrix:
\begin{knitrout}
\definecolor{shadecolor}{rgb}{0.969, 0.969, 0.969}\color{fgcolor}\begin{kframe}
\begin{alltt}
\hlstd{LambdaCorRed} \hlkwb{<-}
  \hlkwd{t}\hlstd{(rotMatCorRed)} \hlopt{%*%} \hlstd{Rmat} \hlopt{%*%} \hlstd{rotMatCorRed}
\hlstd{LambdaCorRed}
\end{alltt}
\begin{verbatim}
##          [,1]
## [1,] 1.738241
\end{verbatim}
\end{kframe}
\end{knitrout}

\item Inverse of dimensionally reduced diagonal eigenvalue matrix,
$\boldsymbol{\Lambda}_{\mathrm{red}}^{-1}$
\begin{knitrout}
\definecolor{shadecolor}{rgb}{0.969, 0.969, 0.969}\color{fgcolor}\begin{kframe}
\begin{alltt}
\hlstd{LambdaCorRedInv} \hlkwb{<-}
  \hlkwd{solve}\hlstd{(LambdaCorRed)}
\hlstd{LambdaCorRedInv}
\end{alltt}
\begin{verbatim}
##           [,1]
## [1,] 0.5752941
\end{verbatim}
\end{kframe}
\end{knitrout}

\item Dimensionally reduced principal component loadings
matrix~$\boldsymbol{A}_{\mathrm{red}}$
\begin{knitrout}
\definecolor{shadecolor}{rgb}{0.969, 0.969, 0.969}\color{fgcolor}\begin{kframe}
\begin{alltt}
\hlstd{AmatCorRed} \hlkwb{<-}
  \hlstd{rotMatCorRed} \hlopt{%*%} \hlstd{LambdaCorRed} \hlopt{^} \hlstd{(}\hlnum{1} \hlopt{/} \hlnum{2}\hlstd{)}
\hlstd{AmatCorRed}
\end{alltt}
\begin{verbatim}
##            [,1]
## [1,] -0.8575507
## [2,] -0.8735813
## [3,]  0.4895956
\end{verbatim}
\end{kframe}
\end{knitrout}

\item Dimensionally reduced matrix~$\boldsymbol{F}$,
$\boldsymbol{F}_{\mathrm{red}}$
\begin{knitrout}
\definecolor{shadecolor}{rgb}{0.969, 0.969, 0.969}\color{fgcolor}\begin{kframe}
\begin{alltt}
\hlstd{FmatCorRed} \hlkwb{<-}
  \hlstd{Z} \hlopt{%*%} \hlstd{AmatCorRed} \hlopt{%*%} \hlstd{LambdaCorRedInv}
\hlkwd{dim}\hlstd{(}\hlkwc{x} \hlstd{= FmatCorRed)}
\end{alltt}
\begin{verbatim}
## [1] 187   1
\end{verbatim}
\begin{alltt}
\hlkwd{head}\hlstd{(}\hlkwc{x} \hlstd{= FmatCorRed)}
\end{alltt}
\begin{verbatim}
##            [,1]
## [1,]  1.8362245
## [2,]  2.6100451
## [3,]  0.6264361
## [4,]  0.2097674
## [5,]  0.4219218
## [6,] -1.1094795
\end{verbatim}
\begin{alltt}
\hlstd{(}\hlnum{1} \hlopt{/} \hlstd{(}\hlkwd{nrow}\hlstd{(FmatCorRed)} \hlopt{-} \hlnum{1}\hlstd{))} \hlopt{*} \hlkwd{t}\hlstd{(FmatCorRed)} \hlopt{%*%} \hlstd{FmatCorRed}
\end{alltt}
\begin{verbatim}
##      [,1]
## [1,]    1
\end{verbatim}
\end{kframe}
\end{knitrout}

\end{enumerate}
%

\subsection{Comparison of trivariate example data set with its dimensionally
reduced variant}
The ``$f$-scores'' remaining in the dimensionally reduced
matrix~$\boldsymbol{F}_{\mathrm{red}}$ need to be transformed back to the
\textbf{original orthonormal basis}, and then expressed with respect to the
\textbf{original scales of measurement}. For illustrative purposes, a sample of
the corresponding dimensionally reduced data will be contrasted with the
original ``$z$-scores,'' and with the original measured values for the three
\textbf{variables} height, mass and age.

\subsubsection{Standardised scale of measurement}
$\boldsymbol{Z}_{\mathrm{red}}$ vs $\boldsymbol{Z}$
\begin{knitrout}
\definecolor{shadecolor}{rgb}{0.969, 0.969, 0.969}\color{fgcolor}\begin{kframe}
\begin{alltt}
\hlstd{Zapprox} \hlkwb{<-}
  \hlstd{FmatCorRed} \hlopt{%*%} \hlkwd{t}\hlstd{(AmatCorRed)} \hlopt{%>%}
  \hlstd{magrittr}\hlopt{::}\hlkwd{set_colnames}\hlstd{(}
    \hlkwc{x} \hlstd{= .,}
    \hlkwc{value} \hlstd{=} \hlkwd{c}\hlstd{(}\hlstr{"height_std [1]"}\hlstd{,} \hlstr{"mass_std [1]"}\hlstd{,} \hlstr{"age_std [1]"}\hlstd{))}
\end{alltt}
\end{kframe}
\end{knitrout}
\begin{knitrout}
\definecolor{shadecolor}{rgb}{0.969, 0.969, 0.969}\color{fgcolor}\begin{kframe}
\begin{alltt}
\hlkwd{head}\hlstd{(}\hlkwc{x} \hlstd{= Z)}
\end{alltt}
\begin{verbatim}
##      height_std [1] mass_std [1] age_std [1]
## [1,]     -1.9300562   -0.9889505   1.3740963
## [2,]     -2.5544936   -1.8466045   1.4974016
## [3,]     -0.8060688   -0.0997265   0.6342642
## [4,]     -0.0567439   -0.6627263  -0.5371365
## [5,]     -0.3065189   -1.2888663  -1.3386213
## [6,]      0.9423560    1.4998244   0.3876535
\end{verbatim}
\begin{alltt}
\hlkwd{head}\hlstd{(}\hlkwc{x} \hlstd{= Zapprox)}
\end{alltt}
\begin{verbatim}
##      height_std [1] mass_std [1] age_std [1]
## [1,]     -1.5746556   -1.6040913   0.8990074
## [2,]     -2.2382460   -2.2800865   1.2778665
## [3,]     -0.5372007   -0.5472428   0.3067003
## [4,]     -0.1798862   -0.1832489   0.1027012
## [5,]     -0.3618193   -0.3685830   0.2065711
## [6,]      0.9514349    0.9692205  -0.5431963
\end{verbatim}
\begin{alltt}
\hlkwd{tail}\hlstd{(}\hlkwc{x} \hlstd{= Z)}
\end{alltt}
\begin{verbatim}
##        height_std [1] mass_std [1] age_std [1]
## [182,]      1.3170185    0.4106565  -0.4754839
## [183,]     -0.6811813   -0.4469974  -0.2042121
## [184,]      0.5676935   -0.1839134   0.5109589
## [185,]      2.5658933    0.9683947  -0.8453999
## [186,]     -1.3056188   -1.4046233  -0.5987892
## [187,]      1.3170185    2.2732915  -1.2153159
\end{verbatim}
\begin{alltt}
\hlkwd{tail}\hlstd{(}\hlkwc{x} \hlstd{= Zapprox)}
\end{alltt}
\begin{verbatim}
##        height_std [1] mass_std [1] age_std [1]
## [182,]     0.84901854   0.86488963 -0.48472437
## [183,]    -0.43150557  -0.43957190  0.24635654
## [184,]     0.03749388   0.03819477 -0.02140612
## [185,]     1.70709764   1.73900920 -0.97462164
## [186,]    -1.01309268  -1.03203088  0.57839810
## [187,]     1.83046773   1.86468550 -1.04505648
\end{verbatim}
\end{kframe}
\end{knitrout}
%

\subsubsection{Original scales of measurement}
This requires a \textbf{backward transformation} (de-standardisation) of the
data in $\boldsymbol{Z}_{\mathrm{red}}$ respectively in $\boldsymbol{Z}$ to
the \textbf{original scales of measurement} used for the three variables height,
mass and age:
\begin{knitrout}
\definecolor{shadecolor}{rgb}{0.969, 0.969, 0.969}\color{fgcolor}\begin{kframe}
\begin{alltt}
\hlstd{b} \hlkwb{<-}
  \hlkwd{attr}\hlstd{(}\hlkwc{x} \hlstd{= Z ,} \hlstr{"scaled:scale"}\hlstd{)}
\hlstd{a} \hlkwb{<-}
  \hlkwd{attr}\hlstd{(}\hlkwc{x} \hlstd{= Z ,} \hlstr{"scaled:center"}\hlstd{)}
\hlstd{Xapp_int} \hlkwb{<-}
  \hlstd{Zapprox} \hlopt{*} \hlkwd{rep}\hlstd{(}\hlkwc{x} \hlstd{= b ,} \hlkwc{each} \hlstd{=} \hlkwd{nrow}\hlstd{(Zapprox))} \hlopt{+}
  \hlkwd{rep}\hlstd{(}\hlkwc{x} \hlstd{= a ,} \hlkwc{each} \hlstd{=} \hlkwd{nrow}\hlstd{(Zapprox))}
\hlstd{XapproxCor} \hlkwb{<-}
  \hlkwd{data.frame}\hlstd{(Xapp_int)} \hlopt{%>%}
  \hlstd{magrittr}\hlopt{::}\hlkwd{set_colnames}\hlstd{(}
    \hlkwc{x} \hlstd{= .,}
    \hlkwc{value} \hlstd{=} \hlkwd{c}\hlstd{(}\hlstr{"height [cm]"}\hlstd{,} \hlstr{"mass [kg]"}\hlstd{,} \hlstr{"age [yr]"}\hlstd{))}
\end{alltt}
\end{kframe}
\end{knitrout}

\medskip
\noindent
$\boldsymbol{X}_{\mathrm{red}}$ vs $\boldsymbol{X}$
\begin{knitrout}
\definecolor{shadecolor}{rgb}{0.969, 0.969, 0.969}\color{fgcolor}\begin{kframe}
\begin{alltt}
\hlkwd{head}\hlstd{(}\hlkwc{x} \hlstd{= X[,} \hlnum{1}\hlopt{:}\hlnum{3}\hlstd{])}
\end{alltt}
\begin{verbatim}
##   height [cm] mass [kg] age [yr]
## 1     139.700  36.48581       63
## 2     136.525  31.86484       65
## 3     145.415  41.27687       51
## 4     149.225  38.24348       32
## 5     147.955  34.86988       19
## 6     154.305  49.89512       47
\end{verbatim}
\begin{alltt}
\hlkwd{head}\hlstd{(}\hlkwc{x} \hlstd{= XapproxCor)}
\end{alltt}
\begin{verbatim}
##   height [cm] mass [kg] age [yr]
## 1    141.5071  33.17148 55.29411
## 2    138.1330  29.52927 61.43916
## 3    146.7821  38.86569 45.68695
## 4    148.5989  40.82686 42.37810
## 5    147.6738  39.82830 44.06286
## 6    154.3512  47.03627 31.90171
\end{verbatim}
\begin{alltt}
\hlkwd{tail}\hlstd{(}\hlkwc{x} \hlstd{= X[,} \hlnum{1}\hlopt{:}\hlnum{3}\hlstd{])}
\end{alltt}
\begin{verbatim}
##     height [cm] mass [kg] age [yr]
## 182     156.210  44.02677     33.0
## 183     146.050  39.40581     37.4
## 184     152.400  40.82328     49.0
## 185     162.560  47.03182     27.0
## 186     142.875  34.24620     31.0
## 187     156.210  54.06250     21.0
\end{verbatim}
\begin{alltt}
\hlkwd{tail}\hlstd{(}\hlkwc{x} \hlstd{= XapproxCor)}
\end{alltt}
\begin{verbatim}
##     height [cm] mass [kg] age [yr]
## 182    153.8304  46.47414 32.85012
## 183    147.3195  39.44581 44.70818
## 184    149.7042  42.01998 40.36509
## 185    158.1934  51.18383 24.90404
## 186    144.3624  36.25369 50.09386
## 187    158.8207  51.86096 23.76159
\end{verbatim}
\end{kframe}
\end{knitrout}
%

\subsection{Visualisation of dimensionally reduced data via 3D scatter plot}
\subsubsection{Standardised scale of measurement}
%
\begin{knitrout}
\definecolor{shadecolor}{rgb}{0.969, 0.969, 0.969}\color{fgcolor}\begin{kframe}
\begin{alltt}
\hlstd{fig5} \hlkwb{<-}
  \hlstd{plotly}\hlopt{::}\hlkwd{plot_ly}\hlstd{(}
    \hlkwc{data} \hlstd{= tibble}\hlopt{::}\hlkwd{as_tibble}\hlstd{(}\hlkwc{x} \hlstd{= Zapprox),}
    \hlkwc{type} \hlstd{=} \hlstr{"scatter3d"}\hlstd{,}
    \hlkwc{x} \hlstd{= Zapprox[,} \hlnum{1}\hlstd{],}
    \hlkwc{y} \hlstd{= Zapprox[,} \hlnum{2}\hlstd{],}
    \hlkwc{z} \hlstd{= Zapprox[,} \hlnum{3}\hlstd{],}
    \hlkwc{mode} \hlstd{=} \hlstr{"markers"}\hlstd{,}
    \hlkwc{size} \hlstd{=} \hlnum{1}
  \hlstd{)} \hlopt{%>%}
  \hlstd{plotly}\hlopt{::}\hlkwd{layout}\hlstd{(}\hlkwc{title} \hlstd{=} \hlkwd{paste0}\hlstd{(}\hlstr{"Dimensionally reduced "}\hlstd{,}
                                \hlstr{"standardised data in Zapprox"}\hlstd{),}
                 \hlkwc{scene} \hlstd{=} \hlkwd{list}\hlstd{(}
                   \hlkwc{xaxis} \hlstd{=} \hlkwd{list}\hlstd{(}\hlkwc{title} \hlstd{=} \hlstr{"height std [1]"}\hlstd{),}
                   \hlkwc{yaxis} \hlstd{=} \hlkwd{list}\hlstd{(}\hlkwc{title} \hlstd{=} \hlstr{"mass std [1]"}\hlstd{),}
                   \hlkwc{zaxis} \hlstd{=} \hlkwd{list}\hlstd{(}\hlkwc{title} \hlstd{=} \hlstr{"age std [1]"}\hlstd{)}
                 \hlstd{))}
\hlstd{fig5}
\end{alltt}
\end{kframe}
\end{knitrout}

\medskip
{\centering \includegraphics[width=14cm]{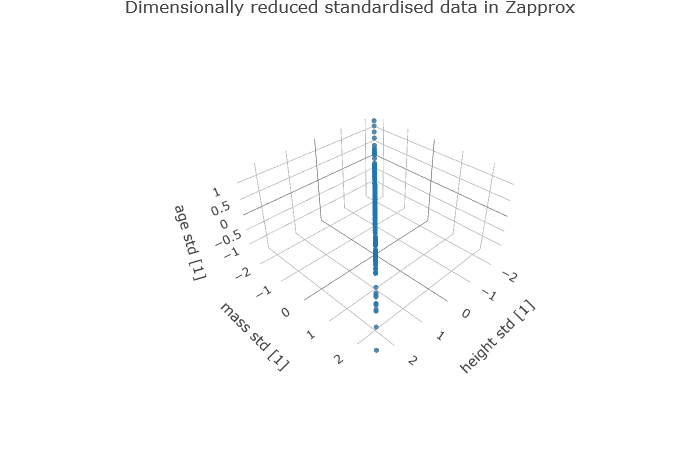}}
%

\subsubsection{Original scales of measurement}
%
\begin{knitrout}
\definecolor{shadecolor}{rgb}{0.969, 0.969, 0.969}\color{fgcolor}\begin{kframe}
\begin{alltt}
\hlstd{fig6} \hlkwb{<-}
  \hlstd{plotly}\hlopt{::}\hlkwd{plot_ly}\hlstd{(}
    \hlkwc{data} \hlstd{= tibble}\hlopt{::}\hlkwd{as_tibble}\hlstd{(}\hlkwc{x} \hlstd{= XapproxCor),}
    \hlkwc{type} \hlstd{=} \hlstr{"scatter3d"}\hlstd{,}
    \hlkwc{x} \hlstd{= XapproxCor[,} \hlnum{1}\hlstd{],}
    \hlkwc{y} \hlstd{= XapproxCor[,} \hlnum{2}\hlstd{],}
    \hlkwc{z} \hlstd{= XapproxCor[,} \hlnum{3}\hlstd{],}
    \hlkwc{mode} \hlstd{=} \hlstr{"markers"}\hlstd{,}
    \hlkwc{size} \hlstd{=} \hlnum{1}
  \hlstd{)} \hlopt{%>%}
  \hlstd{plotly}\hlopt{::}\hlkwd{layout}\hlstd{(}\hlkwc{title} \hlstd{=} \hlkwd{paste0}\hlstd{(}\hlstr{"Dimensionally reduced data "}\hlstd{,}
                                \hlstr{"in XapproxCor"}\hlstd{),}
                 \hlkwc{scene} \hlstd{=} \hlkwd{list}\hlstd{(}
                   \hlkwc{xaxis} \hlstd{=} \hlkwd{list}\hlstd{(}\hlkwc{title} \hlstd{=} \hlstr{"height [cm]"}\hlstd{),}
                   \hlkwc{yaxis} \hlstd{=} \hlkwd{list}\hlstd{(}\hlkwc{title} \hlstd{=} \hlstr{"mass [kg]"}\hlstd{),}
                   \hlkwc{zaxis} \hlstd{=} \hlkwd{list}\hlstd{(}\hlkwc{title} \hlstd{=} \hlstr{"age [yr]"}\hlstd{)}
                 \hlstd{))}
\hlstd{fig6}
\end{alltt}
\end{kframe}
\end{knitrout}

\medskip
{\centering \includegraphics[width=14cm]{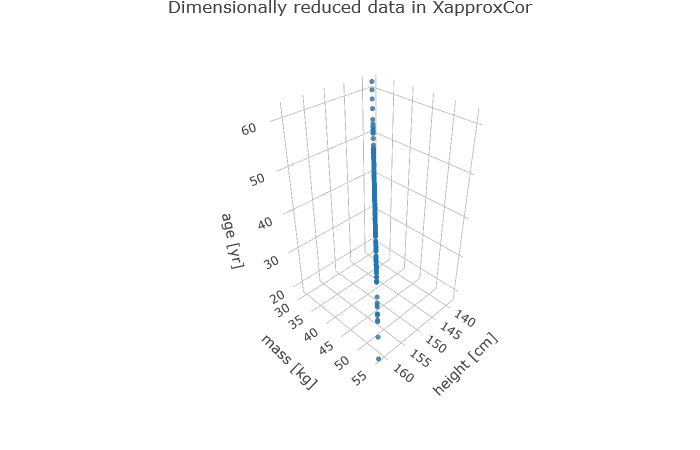}}
%

\subsection{Visualisation of dimensionally reduced data via scatter plot matrix}
\subsubsection{Standardised scale of measurement}
%
\begin{knitrout}
\definecolor{shadecolor}{rgb}{0.969, 0.969, 0.969}\color{fgcolor}\begin{kframe}
\begin{alltt}
\hlstd{GGally}\hlopt{::}\hlkwd{ggpairs}\hlstd{(}\hlkwc{data} \hlstd{= tibble}\hlopt{::}\hlkwd{as_tibble}\hlstd{(}\hlkwc{x} \hlstd{= Zapprox))} \hlopt{+}
  \hlkwd{theme_bw}\hlstd{()}
\end{alltt}
\end{kframe}

{\centering \includegraphics[width=\maxwidth]{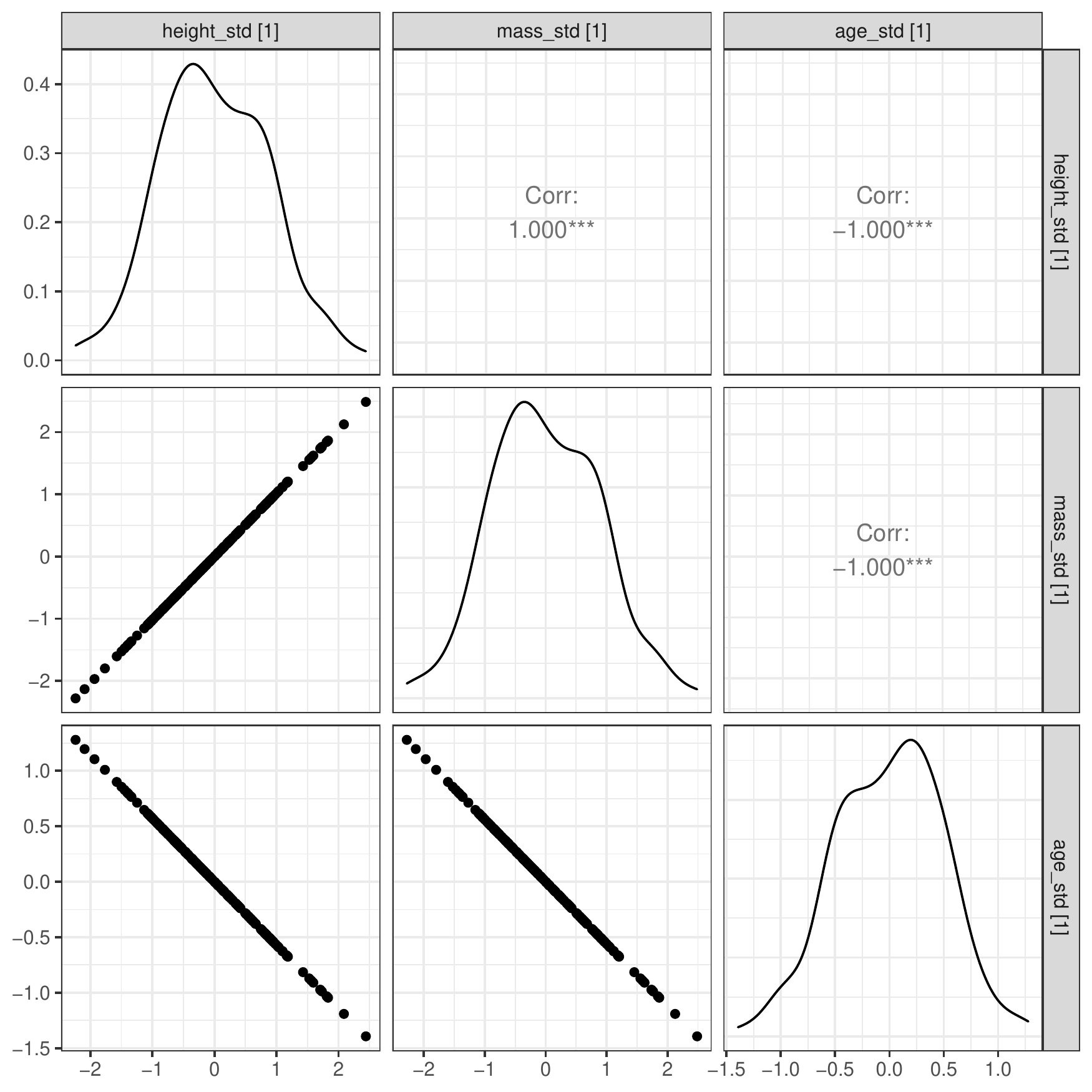} 

}

\end{knitrout}
%

\subsubsection{Original scales of measurement}
%
\begin{knitrout}
\definecolor{shadecolor}{rgb}{0.969, 0.969, 0.969}\color{fgcolor}\begin{kframe}
\begin{alltt}
\hlstd{GGally}\hlopt{::}\hlkwd{ggpairs}\hlstd{(}\hlkwc{data} \hlstd{= tibble}\hlopt{::}\hlkwd{as_tibble}\hlstd{(}\hlkwc{x} \hlstd{= XapproxCor))} \hlopt{+}
  \hlkwd{theme_bw}\hlstd{()}
\end{alltt}
\end{kframe}

{\centering \includegraphics[width=\maxwidth]{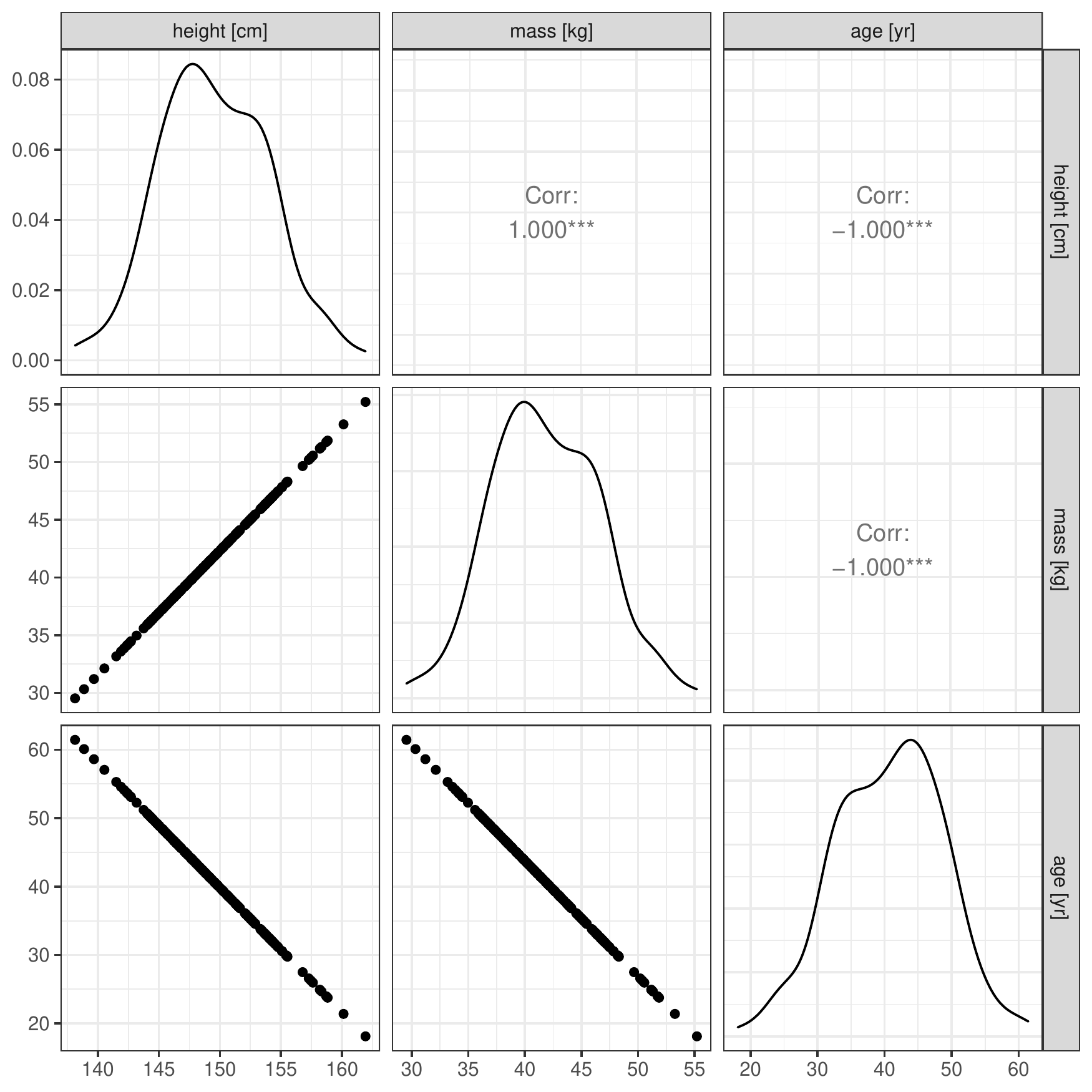} 

}

\end{knitrout}
%

\section{Conclusion}
\lb{sec:concl}
In the example considered, the \textbf{dimensional reduction} performed results
in an \textbf{extremal case}: the \textbf{trivariate data set} in
$\boldsymbol{X}$ was effectively reduced to a univariate data set, which can
explain $57.94~\%$ of the total variance of the original data set.
The dimensionally reduced data set exhibits the maximal (minimal) possible
values for the bivariate correlations between any pair of the three original
variables height, mass and age.

\section*{Acknowledgements}
Constructive comments by Jana Orthey and Laurens van der Woude have helped to
focus this tutorial on the specific needs of the targeted audience.



\end{document}